\documentclass[useAMS,usenatbib]{mn2e}
\usepackage{epsfig}
\usepackage{amsmath}
\def\Pm{\mbox{P}_M}
\def\Rm{\mbox{R}_M}

\def\Rey{\mbox{Re}}
\def\Bmean{{\overline {\bf B}}}
\def\Emean{{\overline {\bf E}}}
\def\Jmean{{\overline {\bf J}}}
\def\Vmean{{\overline {\bf V}}}
\def\Amean{{\overline {\bf A}}}
\def\meanhel{ \langle{\overline {\bf A}}\cdot{\overline {\bf B}}\rangle}
\def\rmsB{ \langle{\overline {\bf B}^2}\rangle}
\def\meancurhel{\langle{\overline {\bf J}}\cdot{\overline {\bf B}}\rangle}
\def\fluchel{ \langle\overline{{\bf a}\cdot{\bf b}}\rangle}
\def\fluccurhel{ \langle\overline{{\bf j}\cdot{\bf b}}\rangle}
\def\urms{{ u_{rms}}}
\def\Brms{{B_{rms}}}
\def\kf{{k_f}}

\newcommand{\be}{\begin{equation}}
\newcommand{\ee}{\end{equation}}

\title[Numerical simulations of decaying helical fields]
{Resilience of helical fields to turbulent diffusion II: direct numerical simulations}
\author[]{Pallavi Bhat$^1$\thanks{
palvi@iucaa.ernet.in}, Eric G. Blackman$^2$\thanks{blackman@pas.rochester.edu} and Kandaswamy Subramanian$^1$\thanks{kandu@iucaa.ernet.in}\\
$^{1}$IUCAA, Post Bag 4, Ganeshkhind, Pune 411007, India.\\
$^{2}$Department of Physics and Astronomy, University of Rochester, Rochester, NY14618, USA}
\begin{document}
%\date{Accepted 1988 December 15. Received 1988 December 14; in original form 1988 October 11}

\pagerange{\pageref{firstpage}--\pageref{lastpage}} \pubyear{2012}

\maketitle

\label{firstpage}

\begin{abstract}
The recent  study of Blackman and Subramanian (Paper I)
 indicates that large scale helical magnetic fields
are resilient to  turbulent diffusion in the sense that helical fields stronger than a critical value, 
decay on slow (resistively mediated), rather than fast ($\sim$ turbulent) time scales. 
This gives more credence to potential  fossil field  origin models of the magnetic fields in stars, galaxies and 
compact objects.   Here we  analyze a suite of direct numerical simulations (DNS)
of decaying large scale helical magnetic fields in the presence of 
non-helical turbulence to further study the physics of helical field decay. 
We study two separate cases:  
(1) the initial field is large enough to decay resistively, is tracked until
it transitions to decay fast, and the critical large scale helical field at 
that transition is sought; 
(2)  the case of Paper I, wherein there is a critical initial helical field strength below which the field undergoes fast decay right from the beginning.   
For case (1), the initial decay rate  in the slow regime is on an average
about twice that of a purely resistive decay and both simulations and solutions 
of the two scale model (from Paper 1), 
reveal  that the transition energy, $E_{c1}$,
is independent of the scale of the turbulent forcing, within a small range of $\Rm$.
We also find that the kinetic alpha, $\alpha_K$, is subdominant to magnetic alpha, $\alpha_M$, in the DNS,
justifying an assumption in the two scale model.
For case (2), we show more comprehensively than in Paper I,  how the two scale theory predicts that 
large scale helical energy at the transition is $E_{c2} = (k_1/\kf)^2 M_{eq}$, where $k_1$ 
and $\kf$ are the large scale and small turbulent forcing scale respectively 
and $M_{eq}$ is the equipartition magnetic energy.  
The DNS in this case agree qualitatively with the two scale model
but the $R_M$ currently achievable,
is too small to satisfy a condition $3/R_M << (k_1/k_f)^2$, necessary to robustly reveal the transition, $E_{c2}$.  
The fact that two scale  theory and DNS agree wherever they can be compared and also 
the two scale theory predicts the transition of case (1) gives us some confidence that 
 $E_{c2}$ of  Paper I  should be identifiable at higher $R_M$ in DNS as well.   
\end{abstract}

\begin{keywords}
dynamo--(magnetohydrodynamics) MHD--turbulence--galaxies:magnetic fields--stars:magnet
\end{keywords}

\section{Introduction}

Astrophysical systems, such as stars, galaxies and 
even galaxy clusters, are observed to host coherent large scale magnetic fields 
\citep{CKB01,Clarke04,govoni_feretti04,BS05,vogt_ensslin,Fletcher10,Beck12}.
The origin of such cosmic magnetic fields has been a long standing open question.
A popular paradigm is that coherent large scale magnetic fields arise
due to dynamo amplification of small seed fields. An interesting alternative
would be if the field from a previous evolutionary phase has simply been
flux frozen when a star, galaxy or a galaxy cluster was formed.
Astrophysical systems are generally turbulent and
such initial fields could then in principle decay due to
turbulent diffusion.
Indeed due to the above reason, the continued
existence of primordial large scale fields in galaxies 
has been mostly considered to be untenable \citep{RSS88}.
However, if coherent magnetic fields in these astrophysical systems were initially of helical nature, 
and sufficiently strong, \citet{BS13} (henceforth Paper I) argued 
on the basis of magnetic helicity conservation, that they
would be resilient to turbulent diffusion and hence, survive up to the present epoch.

 If sub-equipartition helical fields can avoid   turbulent decay, then 
another practical implication is that if helical fields are observed in a system--such as astrophysical jets--
the observed helical  fields would not  necessarily be indicative of  magnetic energy domination in the system.

Magnetic helicity is a nearly conserved quantity in general astrophysical
context and has been useful in understanding of dynamo saturation,
by leading to the development of the dynamical quenching formalism 
(\citet{KMRS00,FB02,BB02,2002PhRvL..89z5007B,Sub02,BS05} 
and references therein).
Paper I 
used the large and small scale magnetic helicity evolution 
equations in a two scale model, along with the mean field induction equation and the
minimal $\tau$-approximation, to understand the decay of helical large scale fields. 
An intriguing result of their work is that even fields which are initially
of fairly sub-equipartition strength, would undergo a slow resistive decay 
if they are helical.
It is important to check the validity of this simple two scale model and the results of Paper I,  
by comparing with results from direct numerical simulations (DNS) 
of decaying large scale helical magnetic fields in presence of non-helical turbulence. 
This is the main motivation of the current work.

There have been previous DNS studies of decaying helical fields
by \citet{YBG03} motivated by trying to understand the quenching of turbulent diffusion.
Also, \citet{KBJ11} discuss simulations of decaying helical fields in
non-helical turbulence, applied to the cylindrical geometry. These simulations
emphasize the decay of initially strong fields of order equipartition value. On the
other hand, Paper I focused on the situation where the initial field strength
is lowered to smaller and smaller values and a threshold energy, $E_{c2}=(k_1/\kf)^2 M_{eq}$
was shown to set the transition from slow to fast decay.
Here $k_1$ and $\kf$ are the wave numbers associated
with the large scale field and the small turbulent forcing scale, respectively
and $M_{eq}$ is the equipartition energy.
Such a threshold was not evident in earlier work.
We wish to examine here through DNS, the decay of helical field
in more generality and with different sets of initial strength and $\kf$.
One motivation is also to examine if there is indeed a $\kf$ dependent threshold
energy.

We limit our present study to initially
fully helical fields, where the measure of helicity
is defined as the ratio of the helical magnetic energy to the
the total magnetic energy.  As will be evident, there is enough richness 
and subtlety to be understood here, even without considering fractionally helical cases.

\begin{figure}
\epsfig{file=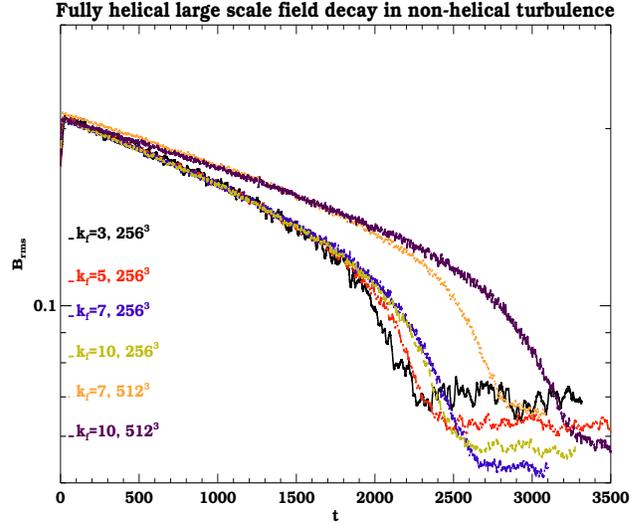, width=0.485\textwidth, height=0.3\textheight}
\caption{The evolution of $B_{rms}$ in helical magnetic field decay
simulations starting with a fully helical 
field of strength 0.2 (superequipartition).
Here, we show the evolution curves for $k_f=3, 5, 7, 10$ from runs of
resolution $256^3 $and also from
two runs with higher resolution of $512^3$ at $k_f=7$ and $10$}
\label{fig:growth}
\end{figure}
In the next section, we discuss  the setup for the simulations and
the quantities to be estimated.
We find that the large scale helical magnetic field
decays in two stages. The first phase is of a slow 
decay, due to only microscopic resistivity. We discuss the slow
regime in detail in Section 2.1.
The second phase comprises of fast decay of the large scale
magnetic field and is discussed in Section 2.2. 
We also estimate the transition point which marks the transition
from slow to fast decay in section 3. There are two kinds of transition points,
arising in two different contexts. One is identified in simulations
of decaying field which start with the same initial field strength
(of equipartition value) and resistivity,
but different $\kf$ (forcing or the turbulent scale).
These show a transition of the evolving field 
from a slow to fast decay regime after decaying to
a critical energy threshold.
The other kind could arise
in simulations of decaying field, where the initial field strength
is decreased until a critical value is reached, below which
the field decays at the fast rate right from the beginning.
The second kind has been emphasized in Paper I and
is discussed in Section 3.2.
In general, throughout the paper, we have juxtaposed the results 
from the simulations with the numerical solutions of the
corresponding two scale model from Paper I.
A discussion of our results and the conclusions are given 
in Section 4.

\section{Simulations of helical large scale field decay}

One of the primary aims of our work is to determine how fast
a helical large scale field decays when subject to turbulent diffusion
by small scale forcing.
We use the
\textsc{Pencil Code}\footnote{http://pencil-code.googlecode.com \citep{B03}}
to simulate the decay of helical large scale fields in the presence of non-helical turbulence.
The fluid is assumed to be isothermal, viscous, electrically conducting
and compressible.
We solve the continuity, Navier-Stokes and induction equations given by,
\begin{eqnarray}
\frac{D~ln\rho}{Dt}&=& -  \nabla \cdot {\bf u},\\
\frac{D~{\bf u}}{Dt}&=& - c_s^2 \nabla~ln \rho + \frac{{\bf J} \times {\bf B}}{\rho} + F_{visc} + f,\\
\frac{\partial {\bf A}}{\partial t}&=& {\bf u} \times {\bf B} + \eta \nabla^2 {\bf A}.
\end{eqnarray}
Here $\rho$ is the density related to the pressure by $ P=\rho c_s^2$, where $c_s$ is speed of sound. 
The operator $D/Dt=\partial/\partial t + {\bf u}\cdot \nabla$ is the lagrangian derivative, where
${\bf u}$ is fluid velocity field.
The induction equation is being expressed in terms of the vector potential, 
${\bf A}$ and ${\bf B}=\nabla \times {\bf A}$, is the magnetic field
${\bf J}= \nabla \times {\bf B}/\mu_0$ is the current density
and $\mu_0$ is the vacuum permeability ($\mu_0=1$ in the DNS).
The viscous force is given by, 
\begin{equation}
F_{visc} = \nu \left[ \nabla^2  {\bf u} + \frac{1}{3} \nabla \cdot\nabla {\bf u} + 2S\cdot\nabla ln \rho\right]
\end{equation}
where,
\begin{equation}
S=\frac{1}{2}\left(\frac{\partial u_i}{\partial x_j}+\frac{\partial u_j}{\partial x_i}-\frac{2}{3}\delta_{ij}\nabla\cdot{\bf u}\right),
\end{equation}
is the traceless rate of strain tensor. The term $f=f({\bf x}, t)$ is responsible 
for turbulent forcing localised in k-space in magnitude and randomly 
changing phase at every time step (see \citet{HBD04} for more details).
These equations are solved in a Cartesian box of a size $l=2\pi$ on a cubic grid with 
$N^3$ mesh points, adopting periodic boundary conditions.
 
The initial magnetic field is a Beltrami field,
${\bf B}=\rm{B}(\rm{sinkz}, \rm{coskz}, 0)$, where the k is set to 1, 
in order to place the field at the largest scale
in the box. Here k corresponding to the box scale is defined as $k=(2\pi)/l$. 
In each run, the helical magnetic field is allowed to
decay under the influence of a non-helical
turbulent forcing. 
We generate the turbulent flow in the box by randomly forcing 
the fluid about an average wavenumber $k_f$, which is much larger than the wavenumber at
which the large scale magnetic field is placed.
The initial velocity field is zero in all the simulations.
We have run a suite of simulations with varying $\kf$ (from 3 to 10),
and initial field strength.
Most of the simulations have a `resolution' of $256^3$,
with 2 higher resolution, $512^3$, runs.
The magnetic and fluid Reynolds numbers throughout this paper are
defined as $\Rm= u_{rms}/\eta k_f$ and $\Rey = u_{rms}/\nu k_f$, respectively,
where $\eta$ and $\nu$ are the resistivity and viscosity of the fluid
and are taken to be equal here and hence, $\Pm=1$.
Table \ref{xxx} gives a list of all the simulations
run towards the study. 
\begin{figure}
\epsfig{file=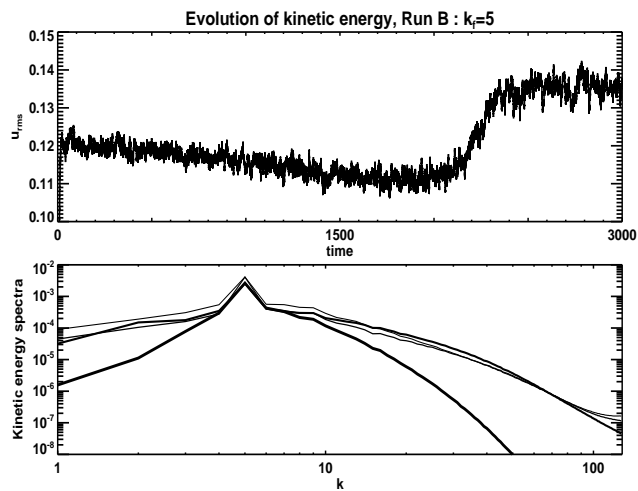, width=0.49\textwidth, height=0.3\textheight}
\caption{The top panel shows the evolution of $u_{rms}$ with time.
It grows to an average initial value of 0.12.
In the bottom panel, we show the spectra of the kinetic energy
at times t=10, 100, 2200 and 2800 for curves with decreasing line thickness respectively.}
\label{figVel}
\end{figure}

Starting with a helical magnetic field,
the rms magnetic field $B_{rms}$, decays exponentially
as shown in Fig.~\ref{fig:growth}, in basically two stages. 
The field decays at a slow rate first and then transitions to a much 
faster rate before finally reaching saturation due to the
floor provided by the fluctuation dynamo (given that all the
simulations have an $\Rm$ which is supercritical enabling the
fluctuation dynamo to operate \citep{Kaz68,HBD04,Schek04,BhatS13}).
In the top panel of Fig.~\ref{figVel}, we show the evolution of $\urms$
for run B with $\kf=5$ (considering this to be the fiducial case).
The kinetic energy decays by less than $10\%$ along with magnetic field in the
first stage. After transition, the magnetic energy decays at 
a fast rate and as a result, the effect of Lorentz forces 
on the velocity field is reduced, thus increasing the $\urms$.
In the bottom panel of Fig.~\ref{figVel}, the corresponding kinetic energy
spectral evolution has been shown at times, t=10, 1000, 2200 and 2800,
with decreasing line thickness.
The peak at $k=5$ corresponds to the constant forcing.

\begin{figure}
\epsfig{file=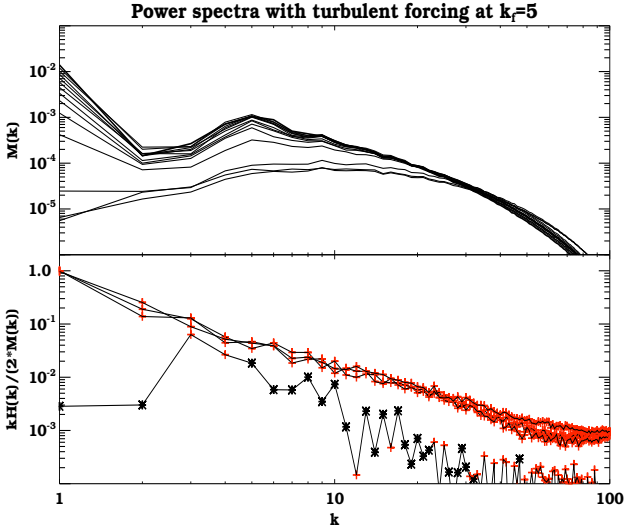, width=0.485\textwidth, height=0.3\textheight}
\caption{The top panel shows the evolving magnetic power spectra, M(k) for run B.
Spectral evolution has been plotted from top to bottom, corresponding to 
times, t=100 to t=2700, with an interval of $\triangle t=200$, between subsequent spectra.
The bottom panel shows the ratio of helical energy, $kH(k)/2$ to the magnetic
energy, $M(k)$. The red 'plus' symbol indicates a positive value, and the black
star indicates a negative value. The four curves from top to bottom are at times 
t=100, 1100, 2100 and 2700}
\label{figmagspec}
\end{figure}

The corresponding evolution of the magnetic energy spectrum, $M(k)$
for run B with $\kf=5$, is shown in the top panel of Fig.~\ref{figmagspec}.
The top spectrum is at t=100 and evolves to the bottom at t=2700, 
with an interval of $\triangle$t=200 between successive spectra.
Initially, the total helicity and the associated helical energy is 
on $k=1$, which is then transferred to smaller scales, on time scale
of few eddy turn over times. In the bottom panel of Fig.~\ref{figmagspec},
we show the fractional helicity spectrum, defined as the ratio
of helical energy, $kH(k)/2$ to magnetic energy, $M(k)$, 
where H(k) and M(k) are the magnetic helicity and energy spectra respectively.
We find that
the helicity on the small scales is of the same sign as that on the 
large scale, as can be seen in the bottom panel of Fig.~\ref{figmagspec}. 
The upper three curves corresponding to times, t=100, 1100, 2100
show that in the large scales, the energy is almost fully helical.
And the fractional helicity in small scales is $< \ 1$,
due to the non-helical energy being constantly pumped at $k=\kf$
(where $\kf > k_1$), due to the 
non helical forcing.
By t=2700, corresponding to the bottom most spectrum of 
highest thickness, the large scale field has almost decayed completely.
And hence, the sign of helicity is fluctuating across all scales.

\begin{table}
\centering
\setlength{\tabcolsep}{3.5pt}
\caption{All the simulations have $\urms=0.12$ and start with 
the fully helical magnetic field of strength $\Brms=0.2$. The runs
are at $\Pm=1$. Also $\overline{\gamma}$ is the average initial decay rate.}
\begin{tabular}{|c|c|c|c|c|c|c|c|}
\hline
\hline
Run & {\small Resolution} & $\kf$ & $\eta\times10^4$ & $\Rm$ & $M_0$ & $\overline{\gamma}$  \\%& $E_{c2}/M_{eq}$ \\ %\hline
\hline
A & $256^3$ & 3 & 2.0   &   200   &  1    & 0.0009   \\
B & $256^3$ & 5 & 2.0   &   120   &  1    & 0.0009   \\
C & $256^3$ & 7 & 2.0   &    86   &  1    & 0.0009   \\
D & $256^3$ & 10 & 2.0  &    60   &  1    & 0.0009   \\
E & $512^3$ & 7 & 1.5   &   120   &  1    & 0.0007   \\
F & $512^3$ & 10 & 1.5  &    80   &  1    & 0.0007   \\
G & $256^3$ & 5 & 2.0   &   120   &  1/5  & 0.003   \\
H & $256^3$ & 5 & 2.0   &   120   &  1/10 & 0.004   \\
I & $256^3$ & 5 & 2.0   &   120   &  1/20 & 0.008   \\
J & $256^3$ & 5 & 2.0   &   120   &  1/25 & 0.010   \\
K & $256^3$ & 5 & 2.0   &   120   &  1/50 & 0.016   \\
\hline
\hline
\label{xxx}
\end{tabular}
%\caption{}
\end{table}

The constant non-helical forcing at $\kf$, generates turbulence
and subsequently facilitates the transfer of helicity and energy from to
$k_1$ to smaller scales.
And then it becomes
imperative to identify the `large' scale field, to be able to analyse the
simulation results. 
We consider contributions from $k=1$ to $k=2$,
to form the large scale field. This seems an appropriate choice
given that the power spectra of various quantities like magnetic
energy and magnetic helicity, have a minimum at k=2 and peak again
at $k=\kf$ as can be seen from Fig.~\ref{figmagspec}.
Consequently, for $k>2$, the spectral energy has 
been considered to be a part of the small scale field.
\footnote{It is difficult to decide an unambiguous scale separation 
between the large and small in general. But in cases such as the 
$\alpha^2$ dynamo, the scale separation can be decided based on the
opposite sign of helicity on the two scales \citep{BS05,AxelPra11}.
In our case as shown in Fig.~\ref{figmagspec}, the helicity on 
both the scales are of the same sign.}
Thus each variable is split into a large scale (mean) and small scale 
(fluctuating) quantity, with an overbar denoting the mean; for
example the magnetic field ${\bf B} = \Bmean + {\bf b}$, where
$\Bmean$ and ${\bf b}$ are respectively the large and small 
scale fields. Further, we will consider volume averages of several
quadratic quantities like magnetic energy density over the whole
simulation box, and denote such  averages by angular brackets $\langle \rangle$.

The three quantities we are mainly interested in are the following. 
First, the total large scale magnetic energy, $\langle\Bmean^2\rangle/2$,
defined as,
\begin{equation}
\frac{\langle\Bmean^2\rangle}{2} = \int_1^2 M(k) \ dk.
\label{bmeaneq}
\end{equation}
Second, large scale helical energy (LSHE), $M_{H}$,
\begin{equation}
\label{LSHE}
M_{H} = \int_1^2 \frac{k H(k)}{2} \ dk.
\end{equation}
This is an important quantity because
the two scale model from Paper I is applied to study the evolution
of LSHE. And then the behaviour of LSHE can be extended
to the total large scale energy, $\langle\Bmean^2\rangle/2$, upto some time scale.
Third, small scale helicity (SSH), $\fluchel$,
\begin{equation}
\fluchel = \int_2^{\infty} H(k)\ dk
\label{ssheq}
\end{equation}
An important understanding derived from the two scale model is 
that the SSH remains in steady state for most of 
the slow decay phase. And the transition from slow
to fast decay is largely governed by the change
in SSH. And hence it is critical to check the nature of
SSH evolution in	 the DNS.

We show the time evolution of $\langle\Bmean^2\rangle/2$,
$M_{H}$ and $\fluchel$
in Fig.~\ref{figdecayk1} for runs A, B,
C and D with a resolution of $256^3$. All these quantities
have been normalised by the equipartition energy 
$M_{eq}=\rho \urms^2/2$, with $\rho\approx 1$ in the simulation units.

These runs have the same $\eta$, but
different $\kf$ and hence different $\Rm$ (see Table~\ref{xxx}).
In Fig.~\ref{figdecay512}, we show the time evolution of these three
quantities in higher resolution ($512^3$) runs E and F.
In the upper panels of Fig.~\ref{figdecayk1} and Fig.~\ref{figdecay512}, 
the evolution of the total large scale energy, 
$\langle\Bmean^2\rangle/2$, is shown in dot-dashed red line 
and the LSHE, $M_H$, is shown in solid black, 
along with the solution for 
$M_H$ from the 
two scale model in blue dashed line.
(The evolution equations for the two scale model are given below).
In the lower panels, the dotted red line shows evolution of SSH, $\fluchel$.
One can observe that the gap between the curve for the two quantities,
$\langle\Bmean^2\rangle/2$ and $M_{H}$ increases once the helicity
has decreased substantially in the fast decay phase. 
Also, the gap becomes more pronounced for smaller
$\kf$ runs due to the smaller scale separation.

The decay rate in a particular decay phase is calculated
by two methods. One is by simply 
fitting an exponential form to the $M_H$(t), given by,
\begin{equation}
M_H(t)=M_{H0} \ e^{-\gamma t}
\label{expeq}
\end{equation}
where  $M_{H0}$ and $\gamma$ are the free parameters.
Note that we retain the code time scale, t, while
plotting the decay curves in the Fig.~\ref{figdecayk1} and Fig.~\ref{figdecay512}.
Here, $t=k_1c_s$, where 
$k_1=1$. %$\tau_{eddy}=t/t_{eddy}$

It can be seen from Fig.~\ref{figdecayk1} and Fig.~\ref{figdecay512},
that slope of LSHE evolution curve is changing continuously.
Thus, the decay rate obtained by the first method, will be an average estimate.
In the second method, we fit for the entire LSHE evolution curve,
using the function,
\begin{equation}
\label{figfitform}
M_H(t) = exp\left( \frac{1}{A+Bt}+\frac{1}{C+Dt} \right)
\end{equation}
where A, B, C and D are free parameters.
\footnote{Other fitting forms were tried, to fit the entire
curve of LSHE evolution. This form provides the best fit
by the method of least squares.}
Then the logarithmic slope of LSHE is derived
from the fit. This gives the decay rate as a function of time.

In both methods of estimating the decay rate, best fits were
decided by the calculation of least squares.
We now discuss the two phases of decay.

\subsection{Slow decay phase}

\begin{figure*}
\epsfig{file=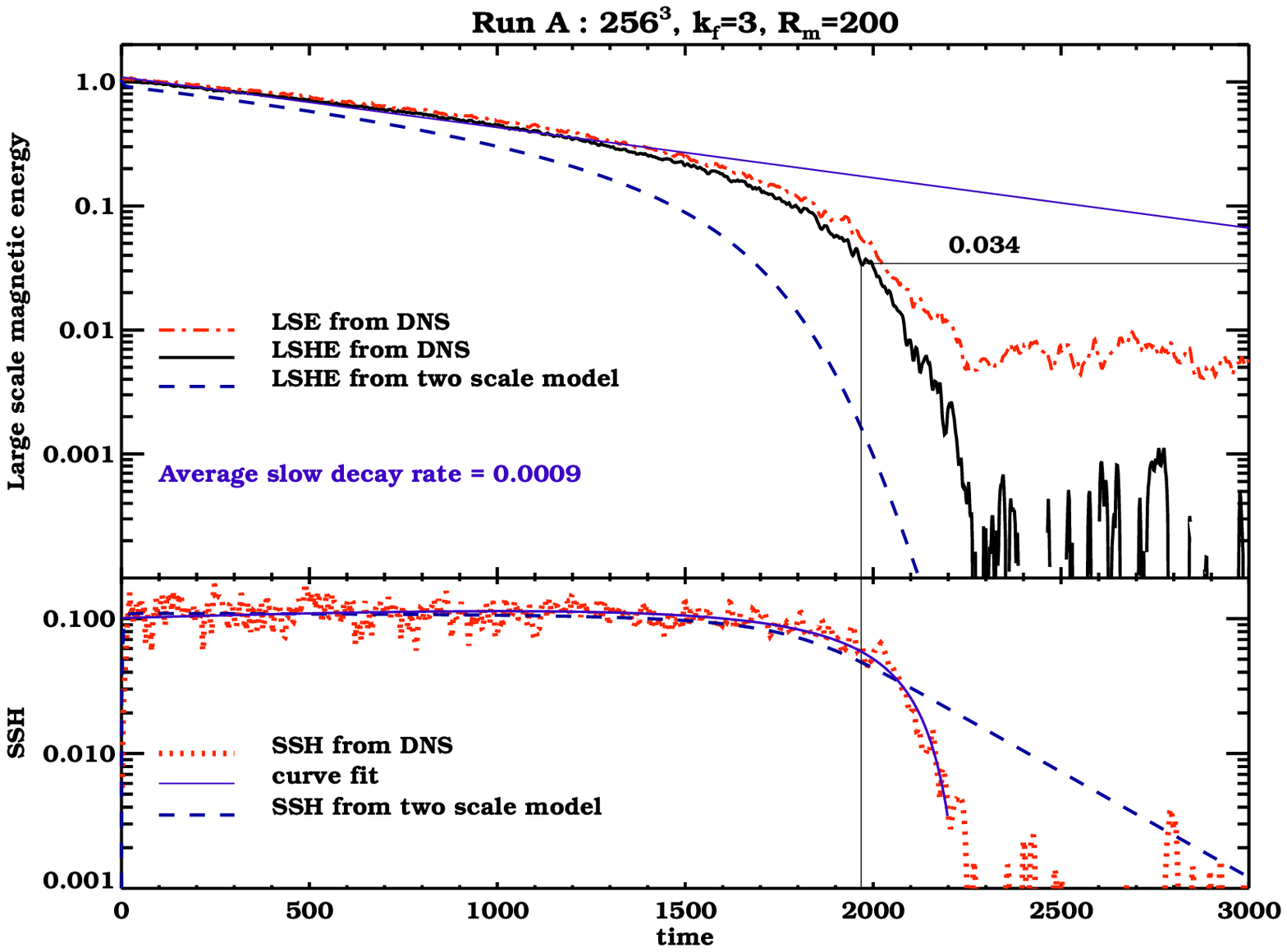, width=0.475\textwidth, height=0.3\textheight}
\epsfig{file=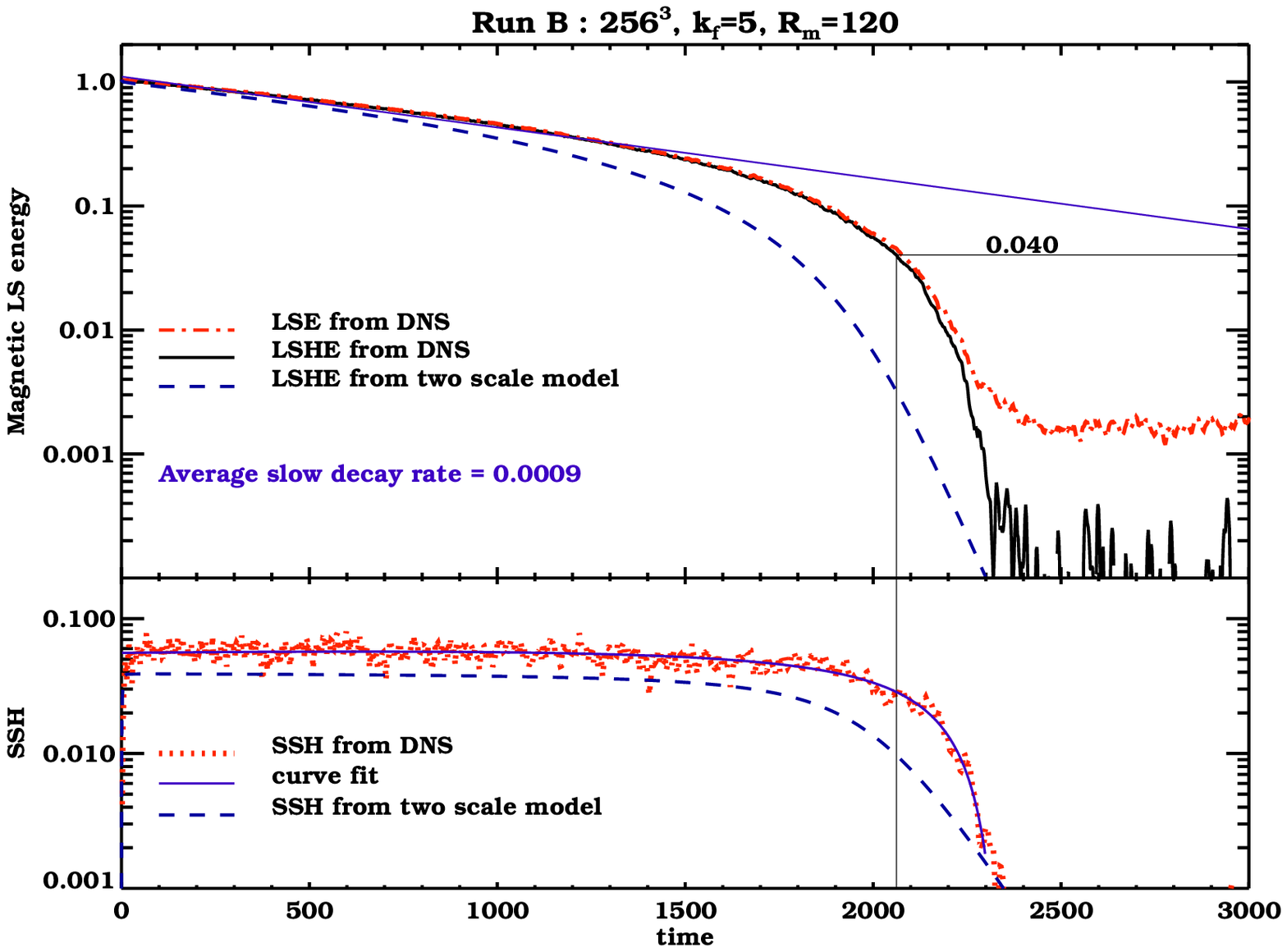, width=0.475\textwidth, height=0.3\textheight}
\epsfig{file=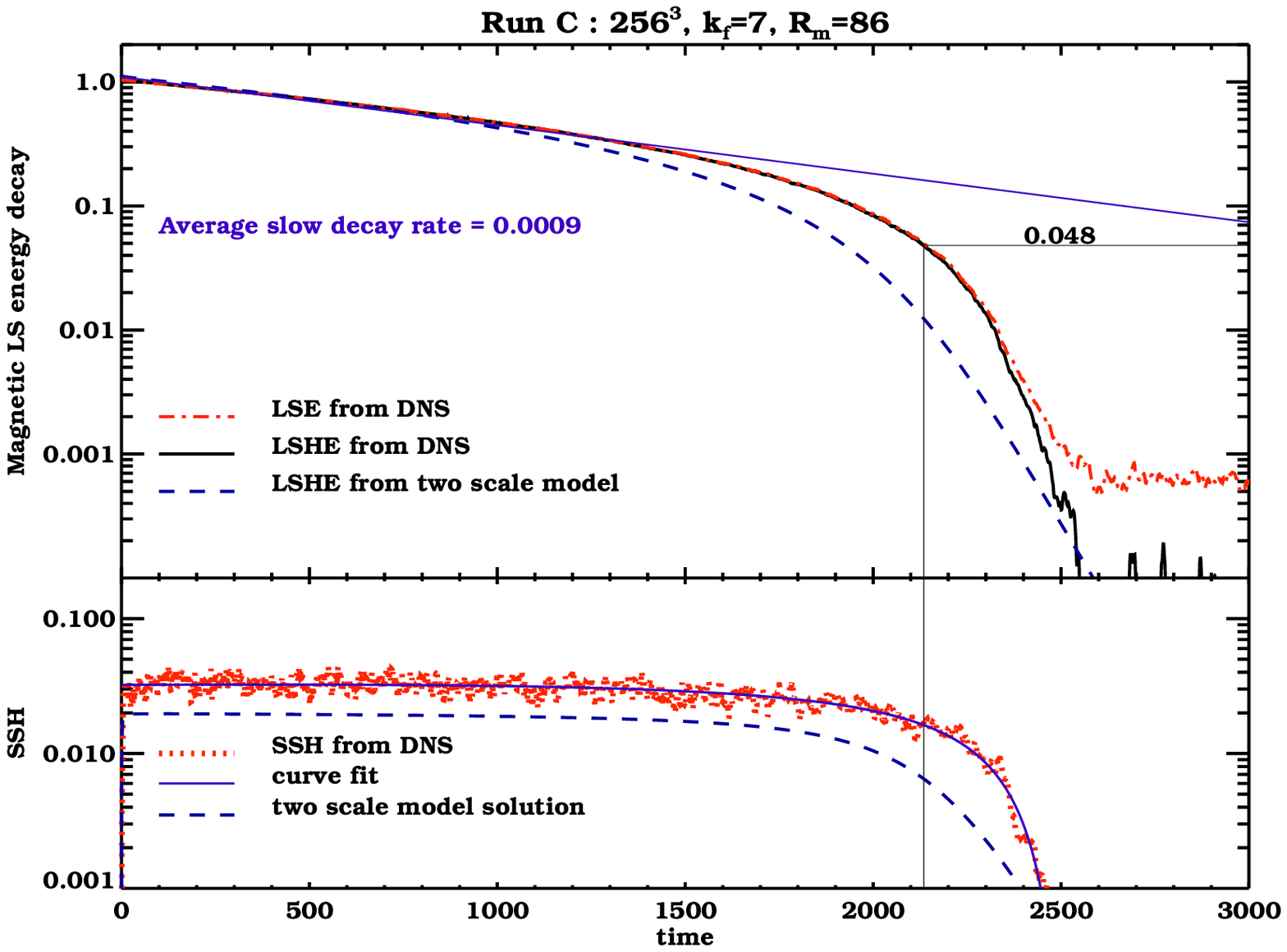, width=0.475\textwidth, height=0.3\textheight} 
\epsfig{file=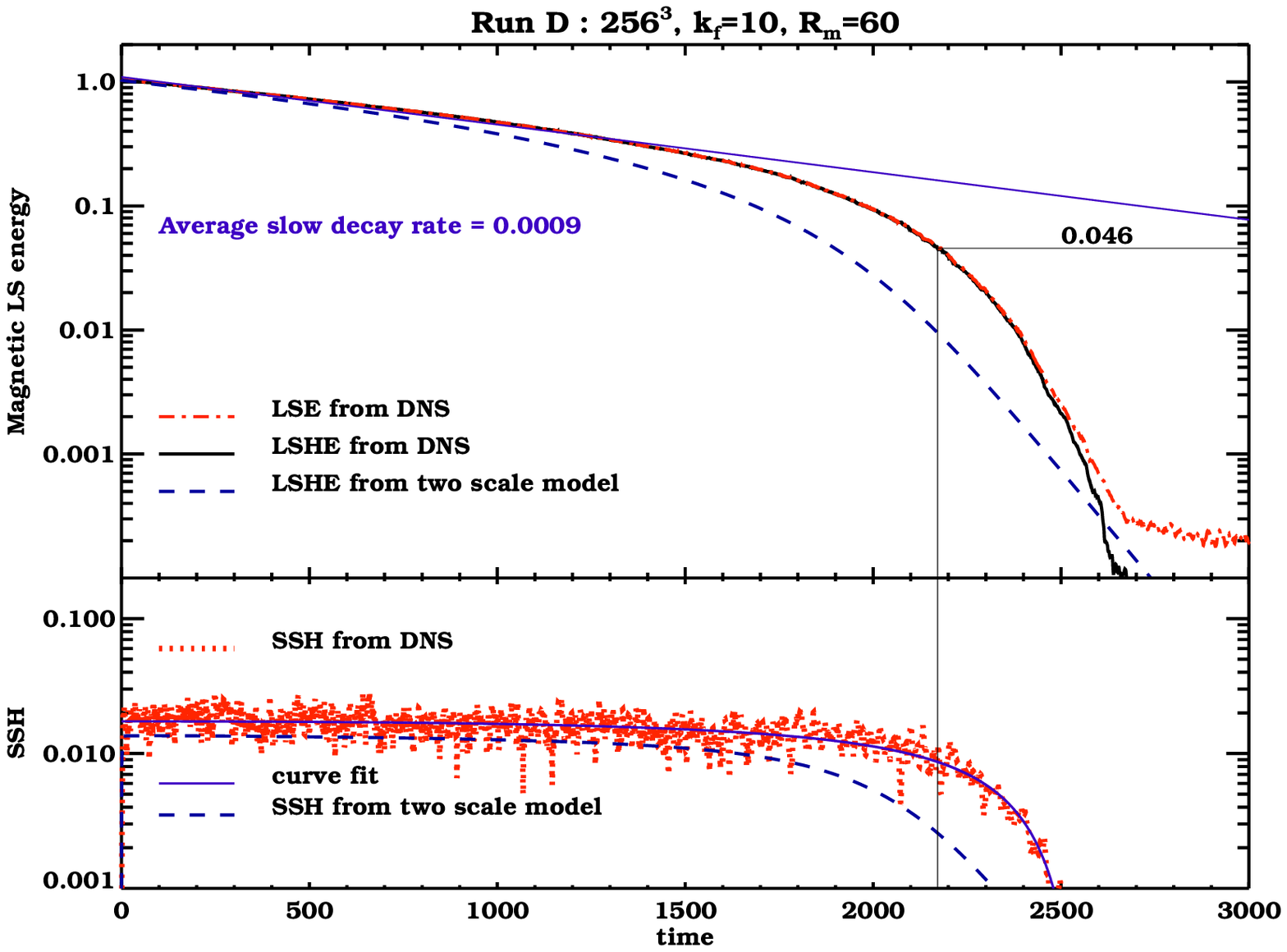, width=0.475\textwidth, height=0.3\textheight}
\caption{We show the evolution of $\langle\Bmean^2\rangle/2$,
$M_{H}$ and $\fluchel$ for runs with $256^3$ resolution at $\kf=3,5,7$ and $10$.
All the quantities are normalised by $M_{eq}$.
The thin vertical line marks the time by when the SSH decreases by $50\%$ of its initial 
steady state value, 
in the bottom panel and intersects the LSHE
curve at the transition energy indicated by the horizontal thin line. The thin 
blue line shows the fit using Eq.~\ref{expeq} to the slow decay phase.}
\label{figdecayk1}
\end{figure*}
The purely resistive decay rate for the large scale magnetic energy
$k_1=1$ mode is given by $2\eta k_1^2=4\times10^{-4}$ (in\
dimensionless units) for runs A to D with $\eta=2\times10^{-4}$.
The slow decay regime for assessing the average decay rate is identified from $t=0$ to $t=t_{slow}$,
where $t=t_{slow}$ is  chosen as an arbitrary time comfortably less than the time, the curve
evolves towards the transition region. 

From the exponential fit to the initial slow decay regime, the average 
decay rate, $\gamma_S\sim 9\times10^{-4}$, is almost twice the purely 
resistive decay rate for $k_1=1$ (where the large scale field resides).
Nevertheless, this $\gamma_S$ is much smaller than the corresponding
turbulent decay rate $\sim 2\eta_t k_1^2 = (2/3)\urms/\kf$.
For example, in the fiducial case of run B, where $\kf=5$ and with $\urms=0.12$,
we have $(2/3)\urms/\kf = 0.016$, which is $\sim$ 18 times larger
than the $\gamma_S$ obtained from DNS.
Notice that the initial field is quite close to the equipartition strength.
This goes to show that helical magnetic field of a sufficiently large
inital strength, decays slowly at a rate which is of the order of the
resistive time-scale and does not decay turbulently as one may have naively expected.
 
In the top panel of Fig.~\ref{figslopeAll}, we show the fit to LSHE
evolution curve for runs A-D, using the form in Eq.~\ref{figfitform}. In the
bottom panel of Fig.~\ref{figslopeAll}, we show the logarithmic slope of 
LSHE from the fit.
On taking the mean of the logarithmic slope values in the resistive
decay phase (from $t=0$ to $t=t_{slow}$), 
we again obtain the average estimate of $\gamma_S\sim 9\times10^{-4}$.
Hence matching with the $\gamma_S$ obtained from the first method.
\begin{figure*}
\epsfig{file=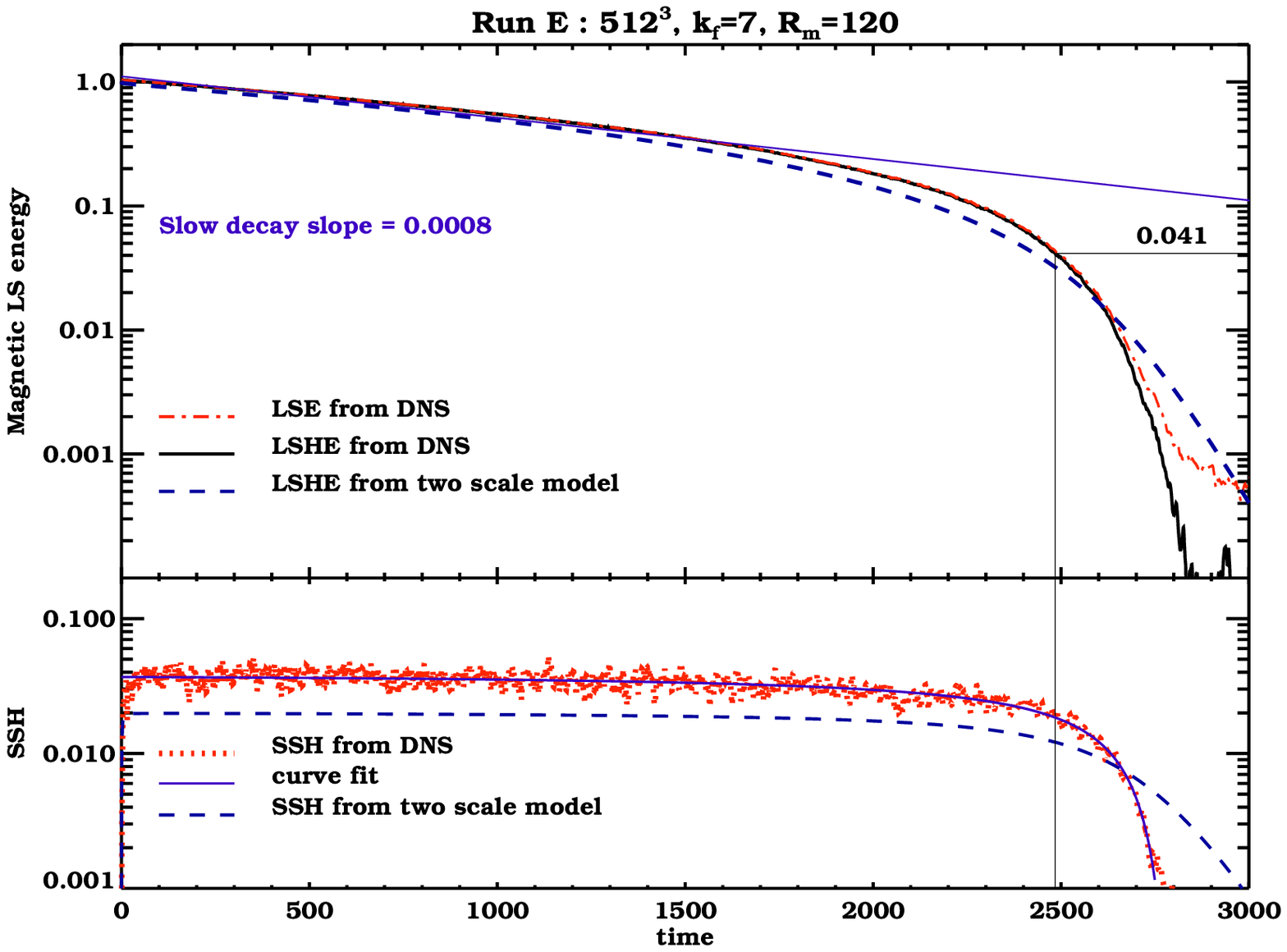, width=0.475\textwidth, height=0.3\textheight}
\epsfig{file=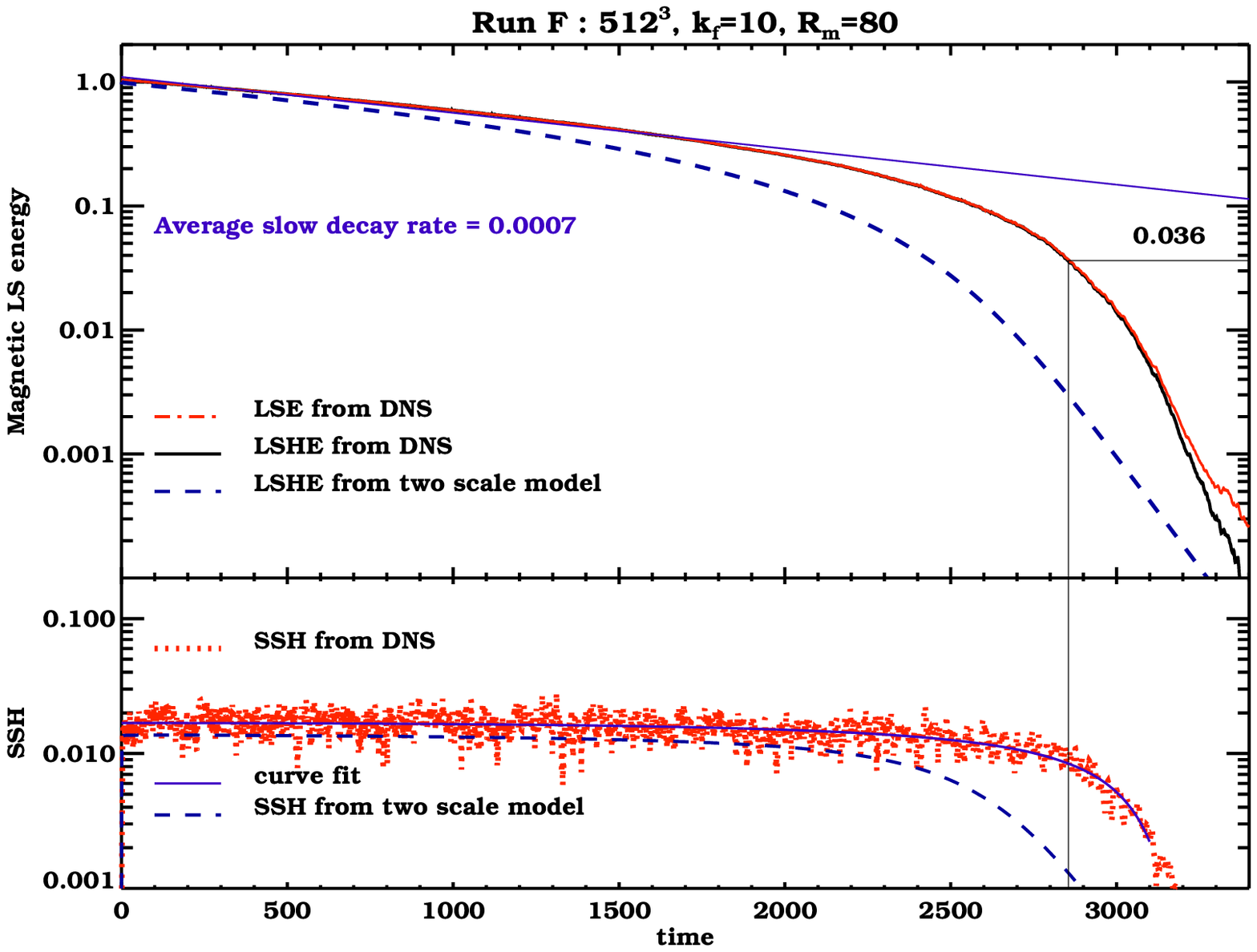, width=0.475\textwidth, height=0.3\textheight}
\caption{We show the evolution of $\langle\Bmean^2\rangle/2$,
$M_{H}$ and $\fluchel$ for runs with $512^3$ resolution at $\kf=7$ and $10$.
All the quantities are normalised by $M_{eq}$.
The thin vertical line marks the time by when the SSH decreases by $50\%$ of its initial 
steady state value, and intersects the LSHE
curve at the transition energy indicated by the horizontal thin line. The thin 
blue line shows the fit using Eq.~\ref{expeq} to the slow decay phase.}
\label{figdecay512}
\end{figure*}
Nonetheless, as can be seen from Fig.~\ref{figslopeAll}, 
the decay rate is continuously changing even in the slow decay phase.
This slowly changing decay rate can be understood by considering the following.

The large scale field in the simulations is almost fully helical.
Hence the large scale field is expected to decay according to the 
equations governing the evolution of magnetic helicity (which is a conserved
quantity in the limit of $\eta \rightarrow 0$).
For periodic or closed domains, evolution of the
total magnetic helicity is given by,
\begin{equation}
\frac{d \langle\overline{{\bf A}\cdot{\bf B}}\rangle}{d t} = -2\eta \langle\overline{{\bf J}\cdot{\bf B}}\rangle
\label{totalheleq}
\end{equation}
All the quantities can be split into a mean (large scale)
and a fluctuating (small scale) component. Accordingly, Eq.~\ref{totalheleq}
can be written as,
\begin{equation}
\label{sepheleq}
\frac{d \meanhel}{d t} + \frac{d \fluchel}{d t} = -2\eta \left( \meancurhel + \fluccurhel \right)
\end{equation}
where cross terms between the large and small scales vanish.
Now, the small scale magnetic helicity is expected to reach steady state
much faster than the large scale magnetic helicity. 
And hence, if $d \fluchel/d t \rightarrow 0$ (as can be seen from Fig.~\ref{figdecayk1}),
then large scale magnetic helicity decays at a rate decided by
both $\meancurhel$ and $\fluccurhel$.

To obtain an equation for large scale helicity, one can use the
mean field induction equation, given as,
\begin {equation}
\frac{\partial \Bmean}{\partial t} = - c\nabla \times \Emean =
\nabla \times \left( \Vmean \times \Bmean + {\bf \mathcal{E}} -
\eta\nabla \times \Bmean \right)
\label{faradayeq}
\end {equation}
where, ${\bf \mathcal{E}}=\overline{{\bf v}\times {\bf b}}$ 
is the electromotive force (or the EMF). Using
the first order smoothing approximation (FOSA) 
\citep{Mof78,KR80} 
or a $\tau$-approximation closure scheme
\citep{PFL76,2002PhRvL..89z5007B,RKR03,BS05},
${\bf \mathcal{E}}$ can be shown to be given by,
\begin{equation}
 {\bf \mathcal{E}} = \left( \alpha_K +\alpha_M \right)\Bmean - \eta_t \Jmean.
\end{equation}
Here the kinetic alpha effect ($\alpha_K$), magnetic alpha effect ($\alpha_M$)
and the turbulent diffusivity ($\eta_t$) are given by,
\begin{equation}
\label{eqalphas}
\alpha_K\simeq - \frac{\tau}{3} \overline{{\bf v}\cdot \boldsymbol{\omega}}, \   
\alpha_M \simeq  \frac{\tau}{3\rho} \overline{{\bf j}\cdot {\bf b}},  \ 
\eta_t \simeq \frac{\tau}{3} \overline{{\bf v}^2 }
\end{equation}
and $\tau$ is the correlation time which can be estimated to be
of order the dynamical or eddy turn over time, $t_{eddy} = 1/(u_{rms}k_f)$.
We uncurl Eq.~\ref{faradayeq} to obtain an equation for the mean vector
potential, $\Amean$. This can be used to obtain the dynamical equation for large scale helicity,
\begin{equation}
\frac{1}{2}\frac{d \meanhel}{d t} = \langle (\alpha_K+\alpha_M) \cdot \Bmean^2 \rangle - \eta_t\meancurhel - \eta\meancurhel
\label{eqlargehel}
\end{equation}
where we have put divergence terms to zero for periodic 
boundary conditions. 
The SSH evolution is obtained by subtracting Eq.~\ref{eqlargehel}
from Eq.~\ref{sepheleq},
\begin{equation}
\frac{1}{2}\frac{d \fluchel}{d t} = - \langle (\alpha_K+\alpha_M) \cdot \Bmean^2 \rangle + \eta_t\meancurhel - \eta\fluccurhel
\label{eqsmallhel}
\end{equation} 
From Eq.~\ref{eqlargehel}, we see that the large scale helical field
would decay due to turbulent diffusion in the absence of the alpha effect.
In our context of forced non-helical turbulence, the kinetic alpha effect
is expected to be negligible. We verify this below, directly from DNS, in section 2.3.
However, $\alpha_M$ could be generated by the
action of turbulent diffusion on a large scale helical field. This 
can be seen explicity in the term, $\eta_t\meancurhel$ in Eq.~\ref{eqsmallhel},
which leads to the generation of $\fluchel$ and hence $\fluccurhel$ (even if
were initially zero), having the same sign as $\meancurhel$. 
Thus, the resulting $\alpha_M$ can in principle, balance the turbulent diffusion
leading to a slow resistive decay of the LSHE. This implicitly constitutes
large scale dynamo action, driven by the small scale current helicity.

\begin{figure}
\epsfig{file=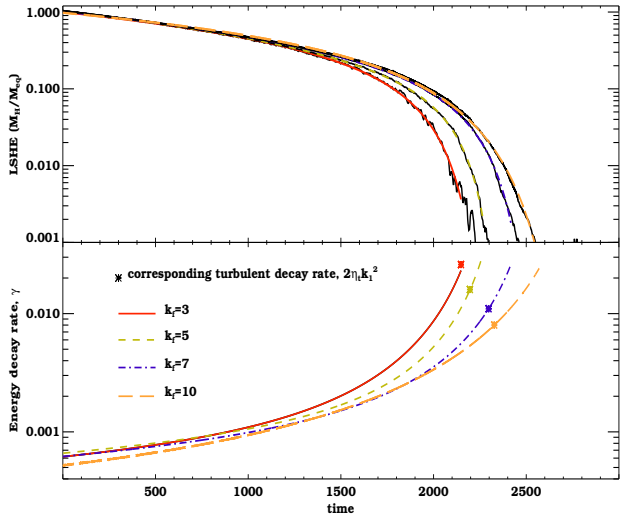, width=0.475\textwidth, height=0.3\textheight}
\caption{Top panel shows the normalised LSHE evolution for runs A-D
along with the fit using the function in Eq.~\ref{figfitform}.
Bottom panel gives the logarithmic derivative using the fit,
i.e.the decay rate evolution for all runs. The symbol, $*$, in the bottom 
panel, marks the value of respective turbulent decay rates.}
\label{figslopeAll}
\end{figure}
We also see from Eq.~\ref{eqsmallhel}, that 
$- \langle \alpha_M \cdot \Bmean^2 \rangle\propto - \kf^2 \fluchel \rmsB $, 
causes rapid damping of the SSH and 
leads to a steady state.
For such a steady small scale helicity, Eq.~\ref{eqsmallhel} can be used
to derive a relation between, $\meancurhel$ and $\fluccurhel$.
We have for $d\fluchel/dt \rightarrow 0$,
\begin {equation}
\label{inter}
0 = -\alpha_M\rmsB + \eta_t\meancurhel - \eta\fluccurhel
\end {equation}
where, we have dropped the kinetic alpha term following 
Paper I. 
And by substituting the expression for $\alpha_M$ 
into Eq.~\ref{inter}, 
we have, 
\begin {equation}
\label{relationJBjb}
\fluccurhel = \frac{\meancurhel}{\frac{\eta}{\eta_t} + \frac{\rmsB/2}{M_{eq}}}
\end {equation}
where, $M_{eq}$ is equal to $\rho u^2_{rms}/2$, in dimensionless units.

In the denominator of 
the RHS of the Eq.~\ref{relationJBjb},
if one considers that $\eta << \eta_t$, and hence negligible,
then with $\rmsB/2M_{eq}=1$ (which is the case initially in the runs A-F), 
we have $\fluccurhel \simeq \meancurhel$.
This implies that from Eq.~\ref{sepheleq}, with $d\fluchel/dt \rightarrow 0$,
the LSHE decays at twice the resistive decay rate. 
This is an average estimate as ($\rmsB/2$) is actually decaying slowly,
and hence, $\fluccurhel/\meancurhel$ will increase over time, increasing the decay rate
as can be seen from the bottom panel of Fig.~\ref{figslopeAll}.
So, in fact the decay rate of LSHE is changing continuously
even in the slow decay phase.  

\begin{figure}
\epsfig{file=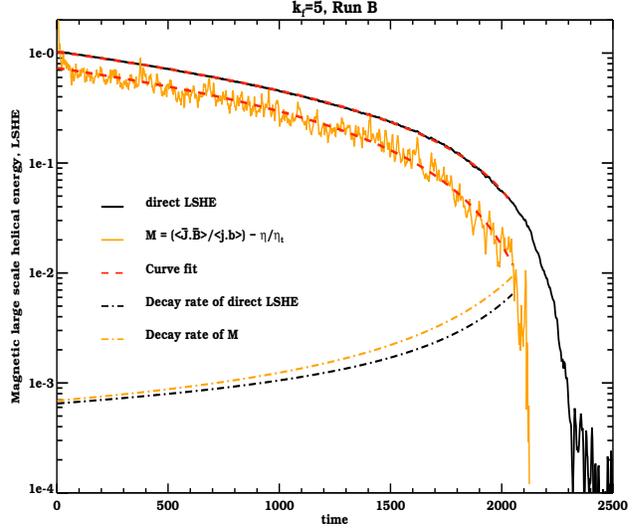, width=0.475\textwidth, height=0.3\textheight}
\caption{The nomalised LSHE evolution curve is overlayed by the predicted noisier
LSHE from the two scale model, as according to 
the Eq.~\ref{relationJBjb}, in the fiducial case of run B.
The dashed curves are the corresponding logarithmic slopes
evaluated using the fit for each of the LSHE curves.}
\label{twoscalematch}
\end{figure}

To further corroborate this continuous change in decay rate from theory, consider the following:
Using the Eq.~\ref{relationJBjb}, we show in Fig.~\ref{twoscalematch}, 
an estimate of the quantity $M=\rmsB/2M_{eq}=\fluccurhel/\meancurhel - {\eta}/{\eta_t}$,
for run B
as a solid yellow line. While LSHE calculated directly using Eq.~\ref{LSHE} is shown in solid black. 
Then, we use the function
in Eq.~\ref{figfitform} to fit for both LSHE and $M$.
Subsequently, we derive the logarithmic slope of the evolution curves
using the fit and have shown them as black and yellow dash-dotted
lines for direct LSHE and $M$, respectively.
While the amplitude of the curve for $M$ is smaller than that of direct LSHE
by $\sim 30\%$, the decay rate evolution predicted by $M$, 
matches closely with that of direct LSHE for most of the resistive decay phase.
And we can see directly from Fig.~\ref{twoscalematch}, that decay rate is increasing
constantly by a small amount for most of the slow decay phase.
The match in the decay rate evolution of $M_H$ with that from the model $M$,
shows that the two scale model is quite useful
for understanding the simulation results in the slow decay phase.

Note that the Fig.~7 of \citet{KBJ11}, shows the
decay of a highly superequipartition field with time.
They find the initial decay rate for the magnetic energy to be $\gamma=-2\eta k^2_1$
(while we obtain the decay rate of $4\eta k^2_1$ for equipartition initial fields).
For $\langle\Bmean^2\rangle/2 >> M_{eq}$, from Eq.~\ref{relationJBjb}, 
$\fluccurhel << \meancurhel$ and hence the large scale field is then
predicted to decay at a purely resistive rate, which is consistent with their finding.

In passing we also note that the correct prediction
of the rate of slow decay, $\gamma_S$ by the two scale model,
which uses the closure relation for $\mathcal{E}$ in the 
Eq.~\ref{eqlargehel}, also lends some credence to such 
mean field closures.

\subsection{Fast decay phase}

We see from Fig.~\ref{figdecayk1} that the LSHE
decays at a much faster rate after it drops below some
critical energy threshold. The fast decay phase is identified from some 
time after the curved transition region, to the time just before
the field saturates, to a level determined by the 
tail of the fluctuation dynamo at large scales.  
(A more precise definition of the transition to
the fast decay regime follows in section 3).
Here the expected decay rate is the turbulent decay rate,
$\gamma_F=2\eta_t k_1^2$, where $\eta_t=u_{rms}/3k_f$.
In the simulations, $\urms=0.12$ and $\eta_t=0.027$ for run A, 
$\eta_t=0.016$ for run B, $\eta_t=0.011$ for run C and $\eta_t=0.008$
for run D. 
The two scale model solutions match with 
expected decay rate of $2\eta_t k_1^2$ and are shown as the 
blue dashed lines in Fig.~\ref{figdecayk1}.
It can be seen from Fig.~\ref{figdecayk1}, that in almost every DNS run,
the slope of the LSHE curve in fast decay phase,
is steeper than that of the two scale model solution.
In fact, we find that the decay rate in the
fast decay phase does not settle to a specific value,
but keeps increasing with time, until
the LS energy has decreased sufficiently to be dominated
by noise.

The top panel of Fig.~\ref{figslopeAll} shows that the fit 
for LSHE evolution curve does not reach an asymptotic slope
at late times. 
The logarithmic slope of the large scale energy derived
from the fit is shown in the bottom panel of Fig.~\ref{figslopeAll} .
It can be seen that for runs B-D, the logarithmic
slope goes to values much larger than the turbulent decay rate.
In all these cases, the large scale and small scale is constituted by a sharp
split at $k=2$. However, this is an imperfect split and 
the effective large scale wave number, $k_1$ could increase
to a higher value, as the large scale field decays. 
Such an increase in the wavenumber for 'large scale' would then increase
the expected turbulent rate. 

Fig.~\ref{figdecayk1} also shows SSH evolution
obtained in the DNS with different $\kf$.
Initially SSH is zero in the DNS,
but rises to a non-zero value due to transfer of helicity
from large to small scales and then stays roughly constant
before decaying at late times. Paper I predicts 
the initial value for the steady state SSH, to be $(k_1/\kf^2) M_{eq}$.
The corresponding two scale solutions are shown as dashed lines in the Fig.~\ref{figdecayk1},
and we see that the steady SSH come close to expected values, 
but are larger in DNS runs B-F.
This shows the limitations of the two scale model
in capturing the whole spectral evolution of the DNS, nevertheless
there is a reasonable agreement with expectations of the two model.
Also, we find that the second slope of SSH, after the steady state phase,
is steeper than the corresponding two scale model slope.
Here again we expect that 
the effective wavenumber
for small scale
field increases from $\kf$ to larger values,
resulting in a faster decay of SSH in the DNS.
Whereas such an
increase would be restricted in the two scale model,
where the small scale is fixed at $\kf$.

\subsection{Effects of kinetic alpha}
Paper I has discussed at length,
the contribution of kinetic alpha, $\alpha_K$ to $\mathcal{E}$.
In the derivation leading to the two scale model, it was
assumed the contribution of $\alpha_K$ to $\mathcal{E}$ is
negligible as compared to $\alpha_M$. 
The kinetic alpha, $\alpha_K$, could be generated due to 
the Lorentz force and then would oppose
the magnetic alpha, $\alpha_M$.
\begin{figure}
\epsfig{file=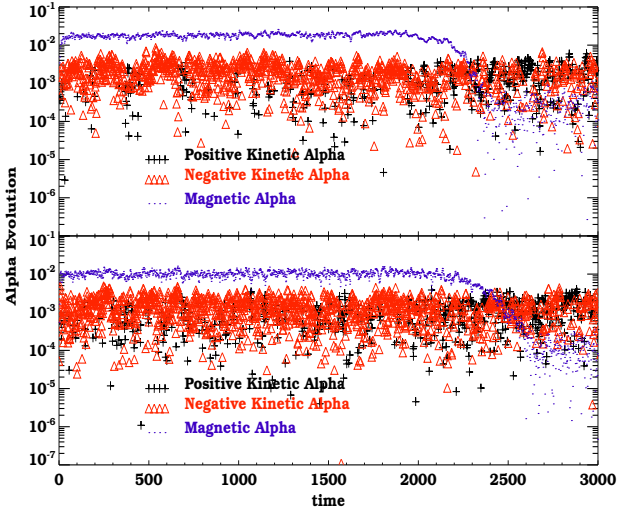, width=0.475\textwidth, height=0.3\textheight}
\caption{The top and bottom panels show $\alpha_K$ and $\alpha_M$ against time
 in the simulations, with $\kf$=5 and $\kf$=10 respectively. }
\label{alpha256b}
\end{figure}
If $\alpha_K$ was significant, then the
large scale helical field would decay much faster
than the resistive decay rate.
It was argued in Paper I that the generated $\alpha_K$
is indeed small.
Nonetheless, it is important to make an estimate of $\alpha_K$
from the DNS and quantify its contribution to the net
$\mathcal{E}$.

In Fig.~\ref{alpha256b}, we show both $\alpha_K$ and $\alpha_M$
estimated from the DNS using Eq.~\ref{eqalphas}.
It can be seen that $\alpha_K$ fluctuates but is mostly negative
and opposite in sign compared to $\alpha_M$ in the slow decay phase.
One also sees from the Fig.~\ref{alpha256b}, that $\alpha_K$ in the slow decay phase is 
found to be a factor of 4-5 smaller than $\alpha_M$ in the case of $\kf=5$ (Run B)
and a factor of $\sim 10$ smaller in the case of $\kf=10$ (Run D).
The contribution of $\alpha_K$ to the EMF is thus considerably smaller than 
$\alpha_M$ and hence subdominant as argued in Paper I.   
In the saturated phase, when all the
magnetic helicity (and hence, $\alpha_M$) has decayed, 
the $\alpha_K$ alternates equally between being positive and negative values.

\begin{figure*}
\epsfig{file=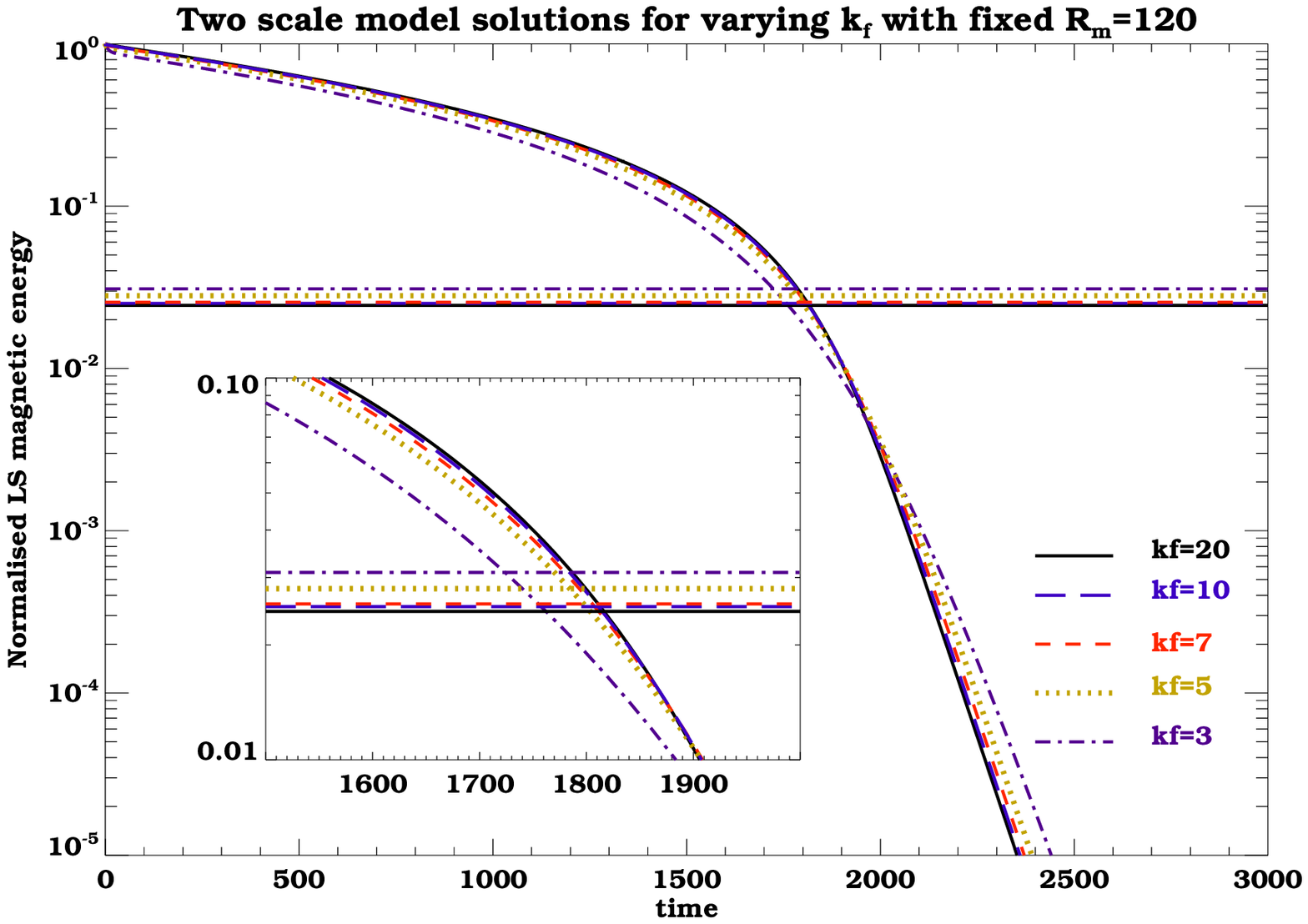, width=0.475\textwidth, height=0.3\textheight}
\epsfig{file=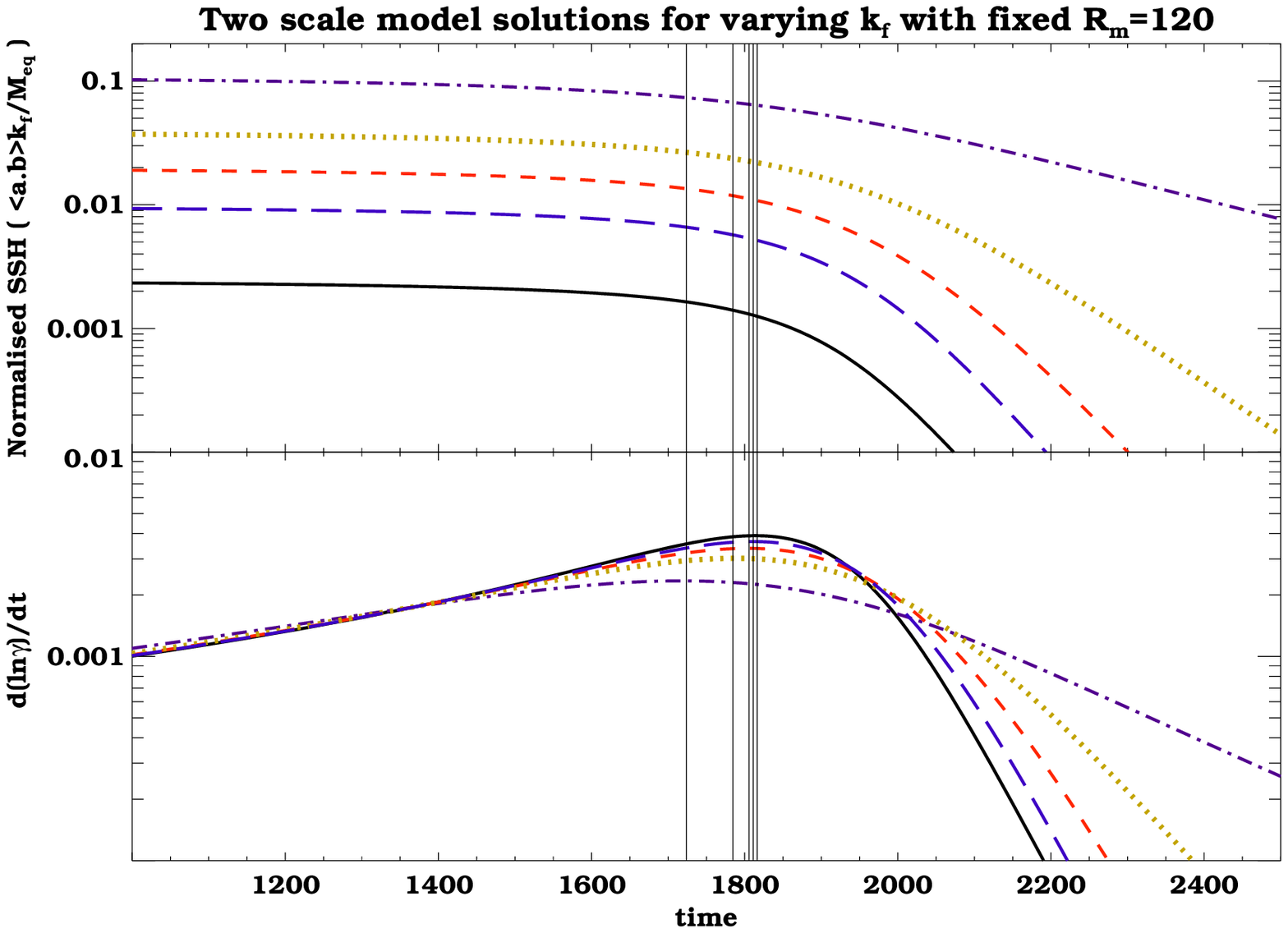, width=0.475\textwidth, height=0.3\textheight}
\caption{ 
For panel in the left, two scale model solutions 
for normalised LSHE is given for varying kf, at 3,5,7,10,20.
Panel in the right shows the corresponding two scale model solutions for normalised SSH.
$\Rm$ is fixed at 120.}
\label{figfixedRm120}
\end{figure*}
\begin{figure*}
\epsfig{file=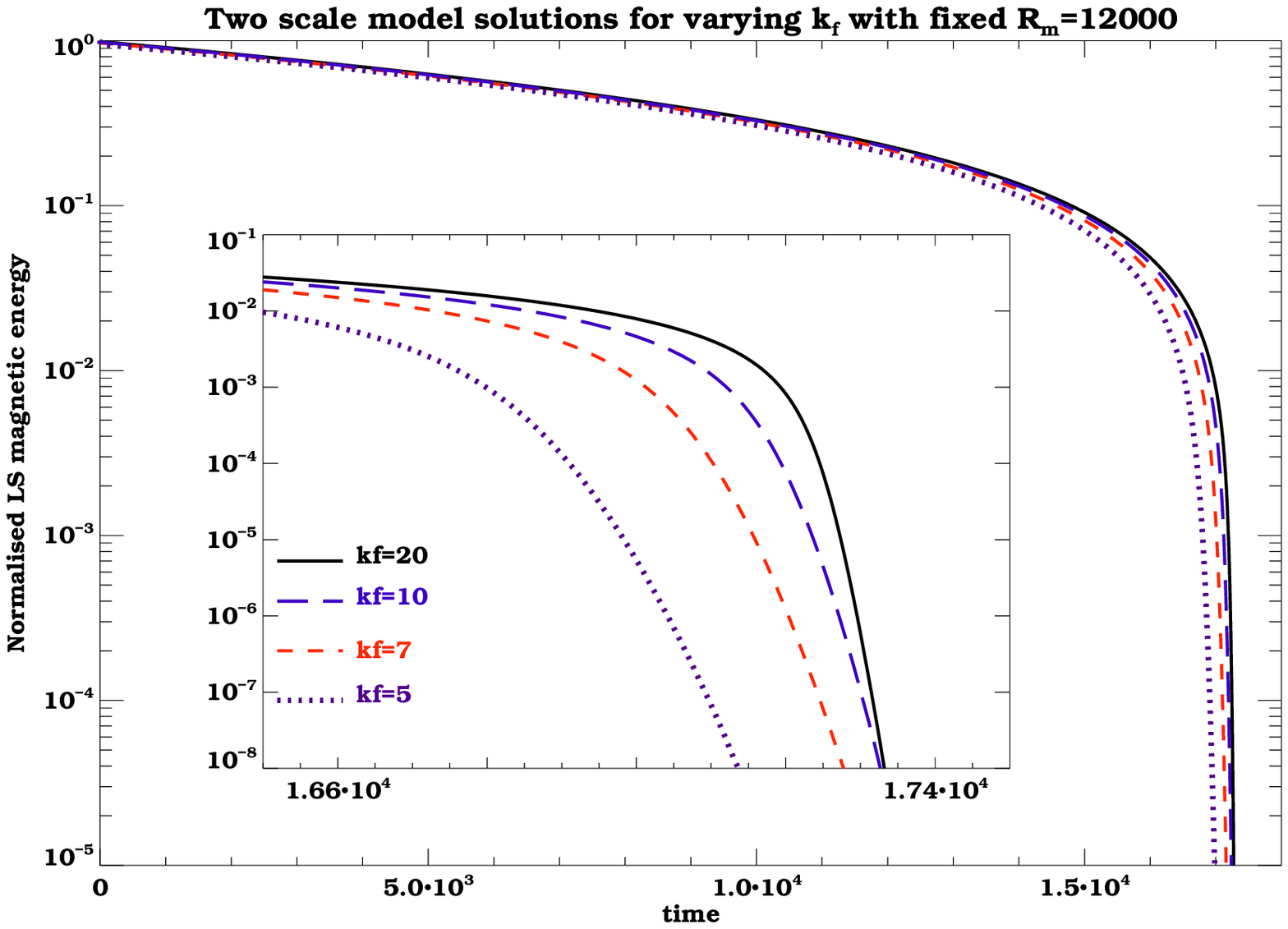, width=0.475\textwidth, height=0.3\textheight}
\epsfig{file=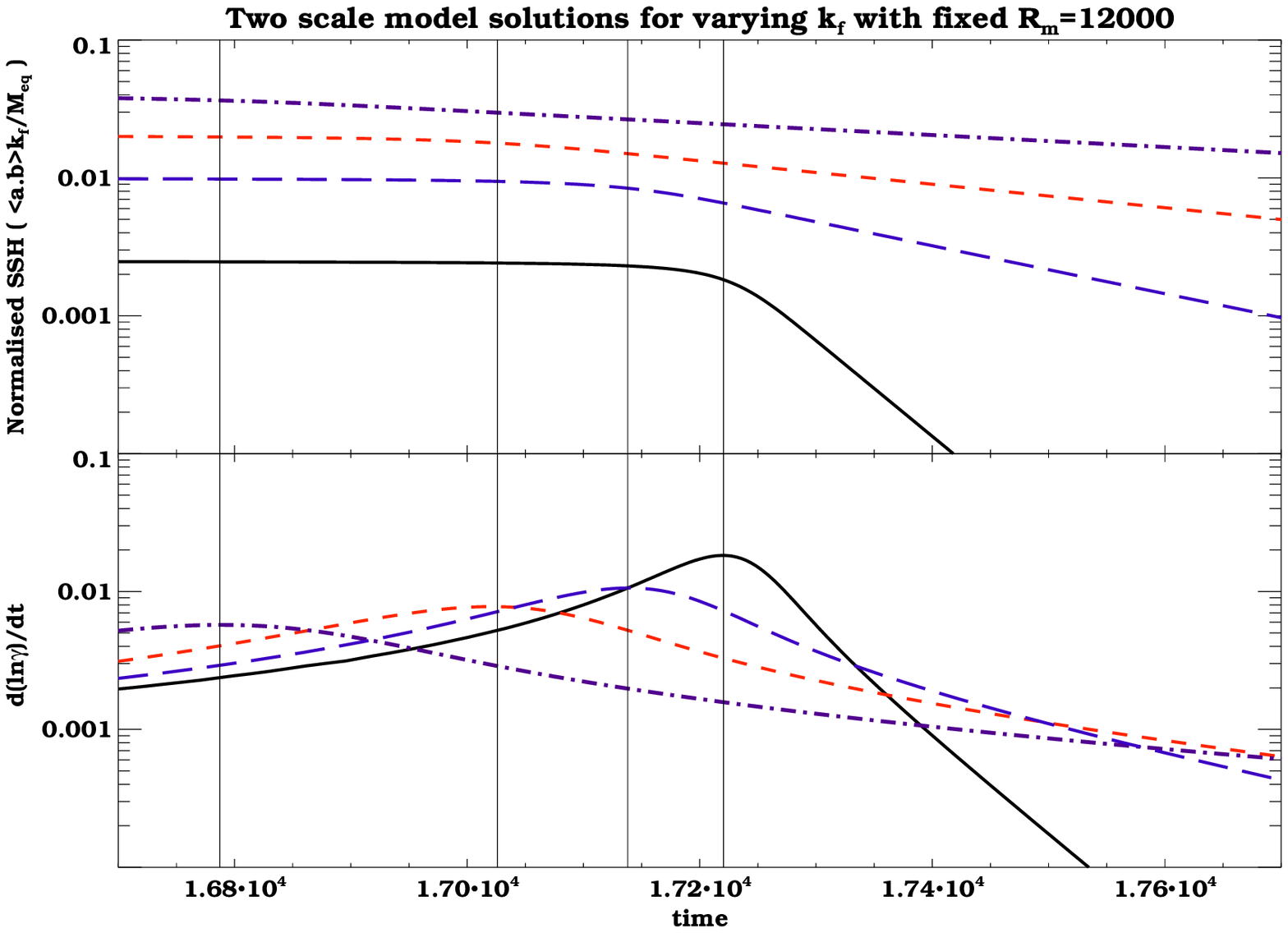, width=0.475\textwidth, height=0.3\textheight}
\caption{
For panel in the left, two scale model solutions 
for normalised LSHE is given for varying kf, at 5,7,10,20.
Panel in the right shows the corresponding two scale model solutions for normalised SSH.
$\Rm$ is fixed at 12000.}
\label{figfixedRm}
\end{figure*}

\section{The transition point}
It is important to identify the threshold below which the
slow decay turns into a fast one, because smaller the
transition energy is with respect to equipartition value,
longer would be the timescale for which the helical large scale
field remains resilient to turbulent diffusion.

We identify two kinds of transition energy, $E_{c1}$ and $E_{c2}$,
arising in two different contexts.
One threshold $E_{c1}$, arises in the context where as the field decays in time, 
it transits from the slow decay phase to the 
fast decay phase after crossing the threshold energy, $E_{c1}$.
This behaviour is what has been examined so far,
in runs A-F, where we started with a field of equipartition strength.
The other context is where one starts with different initial large scale field
strengths. As initial magnetic energy is decreased, below a threshold, $E_{c2}$, the field ceases to start with 
resisitive decay phase and instead decays at a much faster rate right from the beginning.
Paper I argued on the basis of two scale model 
that the latter threshold or critical energy is $\kf$
dependant with $E_{c2} = (k_1/k_f)^2 M_{eq}$.
We will examine both types of transition points. We first focus
on $E_{c1}$ in section 3.1 and later on $E_{c2}$ in Section 3.2.

\subsection{Transition emerging at late times from an initially resistively decaying field}

To identify this transition, 
we first consider the two scale model and then turn to the DNS.
We 
solve numerically Eq.~\ref{eqlargehel} and  Eq.~\ref{eqsmallhel}
of the two scale model,
for different $\kf$, and then plot the time evolution of the large scale magnetic energy. %#### 
To explore the behaviour of the transition point under the variations in $\kf$ alone, 
we keep $\Rm$ and the initial magnetic energy, $M_0$ fixed across different
numerical solutions. 
In the left hand side panel of Fig.~\ref{figfixedRm120}, we show the decay of a fully helical
large scale field with time for different  
$\kf=3,5,7,10,20$. The initial magnetic energy, $M_0=M_{eq}$ 
and the $\Rm$ is fixed to a value of 120 which is comparable to the value in the DNS. 
One immediately notices that all the curves almost coincide.
Note that in these solutions, the $\eta$ is the same, which explains
the same slope in the slow decay regime. And in order to keep the turbulent decay rate, $\sim 2\urms k_1^2/3\kf = 2\urms/3\kf$,
the same, we compensate the increasing $\kf$, by increasing the $\urms$.
This forms the ideal experiment to understand the behaviour
of the transition point as $\kf$ is varied.
From the left panel of Fig.~\ref{figfixedRm120}, the behaviour of the 
transition point is seen to be 
independent of the changing turbulent forcing scale, $\kf$.

To determine the transition point, we have adopted the following method.
The evolving decay rate of the large scale helicity, from the two scale model is
given by,
\begin{equation}
\label{decRate}
\gamma=\frac{1}{2\meanhel}\frac{d\meanhel}{dt}=-\eta_t k_1^2\left(1-\frac{\kf^2\fluchel}{k_1 M_{eq}}\right) - \eta k_1^2
\end{equation}
where, for fully helical fields, $\meancurhel\sim k_1^2\meanhel$ and $\fluccurhel \sim \kf^2\fluchel$.
Also, $\langle\Bmean^2\rangle \sim k_1\meanhel$ and therefore, Eq.~\ref{decRate} also
describes the evolving decay rate of the large scale energy. The decay rate
is expected to be fairly constant in the slow decay phase and sharply increases
during the transition region and then settles to the turbulent decay value.
Thus, the logarithmic slope of $\gamma$ will go through a maximum, when the
decay rate changes the fastest. The point in time when the maximum occurs,
can be then defined as the point of transition and the corresponding
large scale energy is defined to be the transition energy.
In the bottom right panel of Fig.~\ref{figfixedRm120}, we show the 
$d(ln \gamma)/dt$ curves, while
the top right panel shows the evolution of SSH.
Note also that 
the maximum of $d(ln \gamma)/dt$ coincides 
with the point at which the SSH
changes slope, i.e. SSH goes from nearly steady state value to decaying
resistively at $\kf$.
We thus find the transition energy to be $E_{c1}/M_{eq}=$0.031, 0.029, 0.026, 0.025, 0.025
for $\kf=$3, 5, 7, 10, 20 respectively. For completeness, along with the
Eq.~\ref{decRate}, we give here the corresponding equation for the small scale
helical field,
\begin{equation}
\label{decRate2}
\frac{1}{2}\frac{d\ln(\fluchel)}{dt}=-\eta_t \kf^2
\frac{(k_1\meanhel)}{M_{eq}}
\left(1-\frac{k_1 M_{eq}}{\kf^2\fluchel}\right) - \eta k_2^2
\end{equation}

In Fig.~\ref{figfixedRm}, we show similar plots of two scale solutions 
at a much larger $\Rm=12000$, to also test the sensitivity of the results
with respect to changes in $\Rm$. The right panel of the Fig.~\ref{figfixedRm},
shows at the top, evolution of SSH while the bottom panel shows
the evolution of $d(ln \gamma)/dt$. It can be seen from such plots for
both the cases of $\Rm$ = 120 and 12000, that as the $\kf$ increases,
the point at which SSH changes slope, occurs later in time.
And the corresponding large scale energy curve also transitions
later in time. As a result, the transition energy would be
similar across different $\kf$.
The transition energy estimated in this case is $E_{c1}/M_{eq}=$ 0.0011, 0.0009, 0.0009, 0.0008
for $k_f=$ 5, 7, 10, 20 respectively.
We find that the change in the transition energy
from $\Rm=120$ to $\Rm=12000$ is by a factor of 
$\sim 25-30$.
Thus interestingly, $E_{c1}$ seems to scale as $\Rm^{-1/2}$.

Now we turn to the DNS and determine the transition energy of type $E_{c1}$ for the
various runs in Table~\ref{xxx}.
In the case of simulations, we find that the decay rate in fast regime,
is changing with time and and does not settle to a final value
as is the case in the two scale model.
Therefore, we do not find a maximum in the evolving logarithmic slope
of LSHE to be able to determine the transition point. 
Instead, we adopt a slightly different method of estimating the
transition energy in the case of DNS. 
From the right panels of Fig.~\ref{figfixedRm120} and Fig.~\ref{figfixedRm},
we pointed out that $d(ln\gamma)/dt$ is maximum when the SSH begins
to decrease.
Thus it seems plausible to define the transition 
point as the time when the SSH decreases from its initial steady state
value by say, 50\%. From Eq.~\ref{decRate}, it can be seen that the first term
on RHS goes to 0 when $\fluchel = (k_1/\kf^2)M_{eq}$ is in steady state,
and by the time SSH, $\fluchel$ decreases by 50\%, the large scale field is expected
to decay at a rate, of the order of the turbulent decay rate.

We use again the form in Eq.~\ref{figfitform} to determine a fit for the SSH evolution.
Then we estimate the point in time by when SSH decreases by 50\% of its initial steady state value.
The corresponding value of large scale energy is the transition energy.
This method of determining the transition energy is conceptually 
similar to the one used for two scale model as it determines 
the point at which there is a change in the SSH evolution from
steady state (or nearly zero decay rate) to a non-zero decay rate.
This method of determining the transition energy is 
illustrated in the lower panels of Fig.~\ref{figdecayk1} and Fig.~\ref{figdecay512}.
The vertical lines in these figures give the time when
SSH has decreased by $50\%$ of its initial steady state value.
This line intersects the LSHE curve at a transition energy
value indicated by the horizontal line in each upper panel of 
Fig.~\ref{figdecayk1} and Fig.~\ref{figdecay512}.
The transition energy thus determined, 
gives the critical energy as $E_{c1} \sim 0.034 \ M_{eq}$ for run B. 
This is similar to the transition point we obtain from the
corresponding two scale model solution of $E_{c1} \sim 0.029 M_{eq}$.
For the other runs A, C, D, E and F we find the transition energy 
to be $E_{c1}/M_{eq}=0.052, 0.049, 0.048, 0.037$ and $0.034$ respectively,
as can be read from the Fig.~\ref{figdecayk1} and Fig.~\ref{figdecay512}.
This indicates the near universality of the transition point, $E_{c1}$, 
with respect to $\kf$ in a small range of $\Rm$.

\begin{figure*}
\epsfig{file=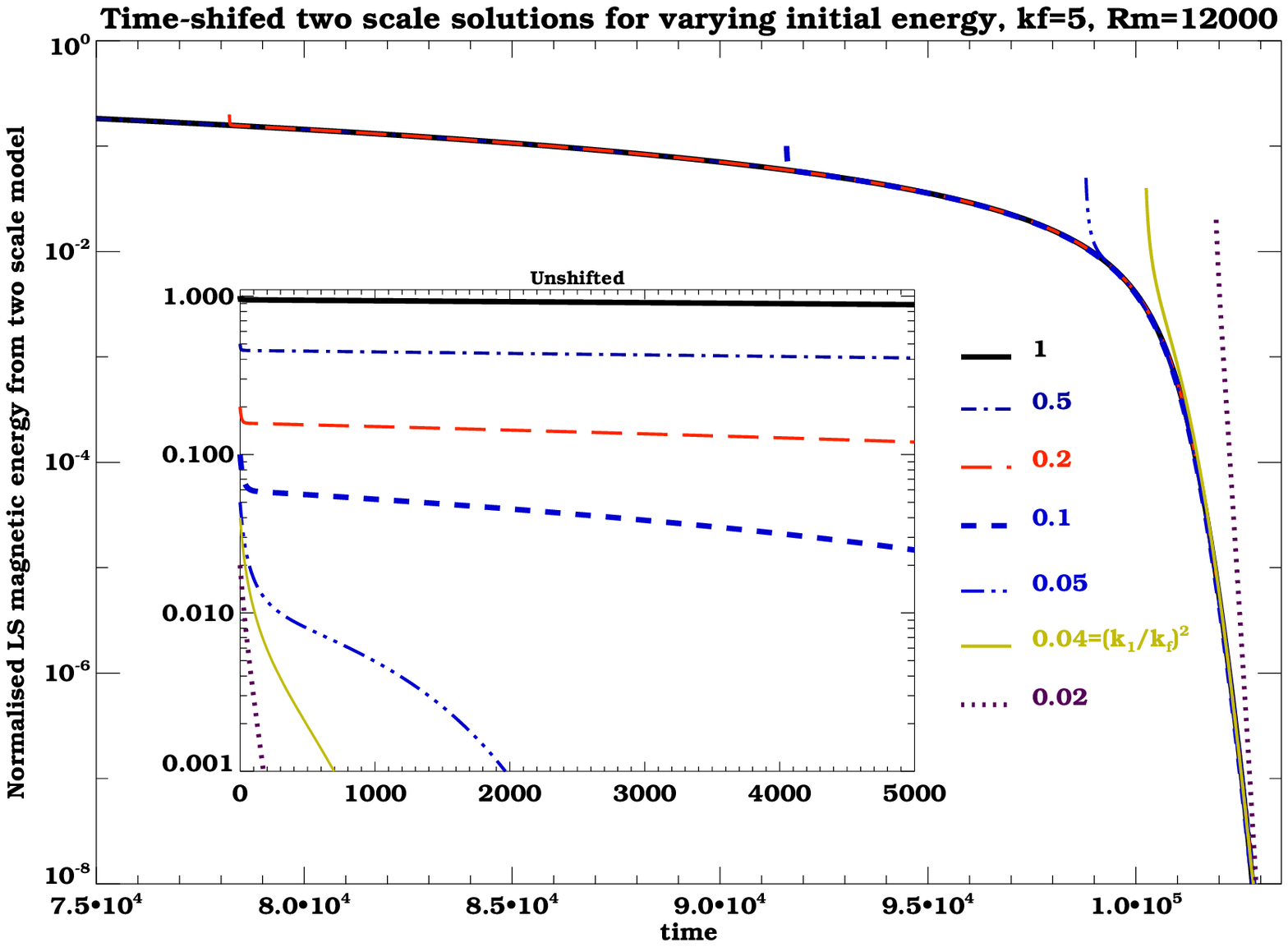, width=0.475\textwidth, height=0.3\textheight}
\epsfig{file=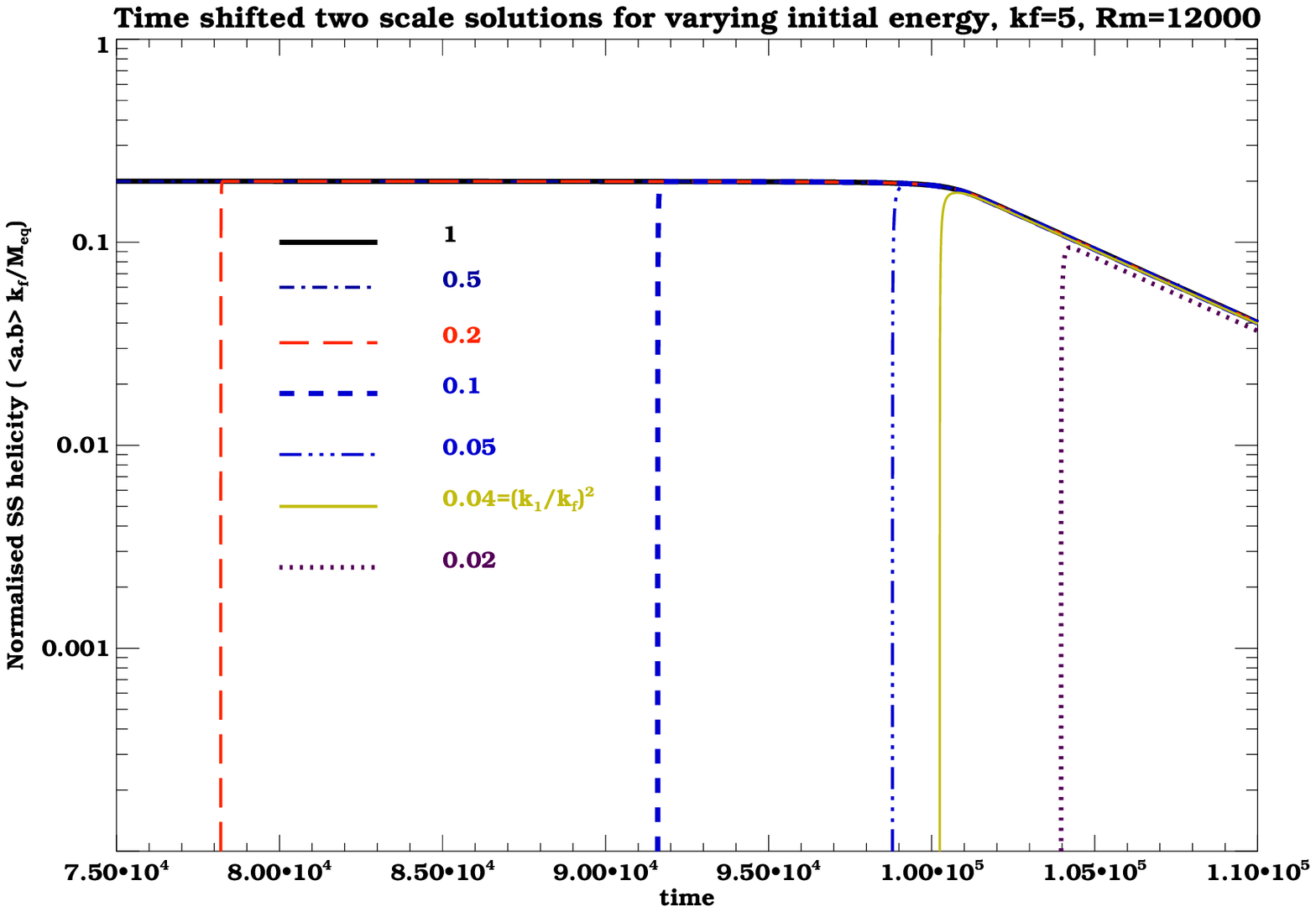, width=0.475\textwidth, height=0.3\textheight}
\caption{The decay curves from two scale model for fully helical large scale magnetic field for 
different initial strengths for $\kf=5$ and $\Rm=12000$ are shown.}
\label{figvaryinit}
\end{figure*}
One can also determine the transition energy from 
the point of intersection between the slopes fit to
the two decay regimes, slow and fast.
The slope intersection method depends on accurate fits for the two 
decay regimes and hence is subject to uncertainity. 
Also, as we see in Fig.~\ref{figslopeAll} that the slopes are not constant 
in any of the decay regimes, and are continually changing, and 
therefore, the determined slope is an approximate average estimate.
The fit in especially the fast decay regime seems highly 
uncertain, depending on the window of time chosen.
Hence we do not pursue this method for determining the transition
energy.

\subsection{Transition identified by varying the initial field strength}

Now we will examine the second kind of transition point, $E_{c2}$. 
This occurs when the initial magnetic field strength is lowered
to a critical point below which the the field decays at a fast decay rate
right from the beginning (i.e. the initial slow decay phase is now absent).
We consider the evolution Eq.~\ref{eqlargehel} for the large scale helicity
and substitute $\fluccurhel$ in $\alpha_M$ (in the emf
$\mathcal{E}$ term) and $\meancurhel$, with the corresponding two scale
approximation of $\kf^2 \fluchel$ and $k_1^2 \meanhel$, respectively.
We then get,
\begin{equation}
\label{LSHeveq}
\frac{1}{2}\frac{\partial \meanhel}{\partial t} =  \frac{1}{3\rho} \kf^2 \fluchel \tau \langle \Bmean^2 \rangle -  \frac{\urms}{3\kf} k_1^2 \meanhel  - \eta k^2_1 \meanhel
\end{equation}
Let us focus on the ideal limit of $\eta\rightarrow0$, for which the total magnetic helicity is conserved
at all times. Then, one can substitute for the small scale helicity in Eq.~\ref{LSHeveq}, 
$\fluchel = \left(\langle\overline{{\bf A}\cdot{\bf B}}\rangle - \meanhel\right)$, 
where the total helicity $\left(\langle\overline{{\bf A}\cdot{\bf B}}\rangle\right)$ is conserved.
Converting all the quantities to a dimensionless form, we get,
\begin{equation}
\frac{dM_1}{d\tilde{t}}+ \frac{2}{3} M_1^2 - \frac{2}{3} M_1 \left( M_0 - (k_1/\kf)^2 \right) = 0
\label{Meq}
\end{equation}
where $M_1 = \langle\Bmean^2\rangle/(2M_{eq})=k_1 \meanhel/(2M_{eq})$, 
$t/\tau = \tilde{t}$ and 
$M_0=M_1(\tilde{t}=0)$ 
is the normalised initial  energy of the large scale helical field.
Eq.~\ref{Meq} can now be solved to give,
\begin{equation}
M_1= \frac{M_0 - (k_1/\kf)^2}{1-\left(\left({(k_1/\kf)^2 e^{(-2\tilde{t}(M_0-(k_1/\kf)^2)/3)}}\right)/{M_0}\right)}
\end{equation}
When $M_0 > (k_1/\kf)^2$, at late times $\tilde{t}\to\infty$, we have,
\begin{equation}
M_1 \to M_0 - (k_1/\kf)^2
\end{equation}
indicating that, the reduction in the field strength is
by a finite amount.
When $M_0 = (k_1/\kf)^2$, Eq.~\ref{Meq} becomes $dM_1/dt = - (2/3)M_1^2$ and hence $M_1 \rightarrow 0$
at late times.

And when $M_0 < (k_1/\kf)^2$, we obtain at late times when 
$\tilde{t}\to\infty$,
\begin{equation}
M_1=M_0\left( 1 - (\kf/k_1)^2 M_0\right) e^{\left({-2\tilde{t}((k_1/\kf)^2-M_0)}/{3}\right)}
\end{equation}
which implies that the large scale field undergoes a rapid decay.
Hence, $(k_1/\kf)^2$ forms a natural
transition point, in the case of large $\Rm$ (or here in the ideal
limit of $\Rm \rightarrow \infty$), which determines
when the LSHE will directly transit to
rapid decay. This was emphasized in Paper I
but without giving the above argument.
The question arises how well this threshold, which holds in 
the ideal limit, obtains for more realistic $\Rm$, both in
the two scale model and the DNS. We first reconsider the
two scale model.

\begin{figure}
\epsfig{file=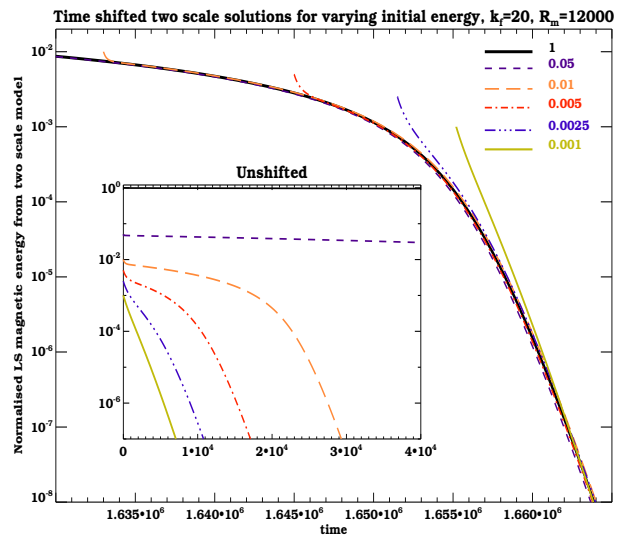, width=0.475\textwidth, height=0.3\textheight}
\caption{The decay curves from two scale model for fully helical large scale magnetic field with
different initial strengths for $\kf=20$ and $\Rm=12000$ are shown.}
\label{figvaryinitkf20}
\end{figure}

The left panel of Fig.~\ref{figvaryinit}, and Fig.~\ref{figvaryinitkf20},
show the decay of fully helical large scale magnetic field 
for a set of decreasing initial field strengths for the two scale model. We have 
adopted $\kf=5$ (Fig.~\ref{figvaryinit}) and $\kf=20$ (Fig.~\ref{figvaryinitkf20}),
both with $\Rm=12000$.
While the inset in the left panels of Fig.~\ref{figvaryinit} and Fig.~\ref{figvaryinitkf20}
show the evolution of large scale energy with decreasing initial strengths,
the main plots show the curves starting with subequipartition strength,
time-shifted to maximally coincide the $M_0=1$ evolution.
The labels in the plot indicate the value of $M_0$.
The thick black curve is the case of $M_0=1$ (we will refer to this as the fiducial curve). 
In the curves beneath that of $M_0=1$, $M_0$ has been decreased to smaller and smaller values.

The right panel of Fig.~\ref{figvaryinit} shows the 
evolution of the SSH. 
For stronger initial fields, SSH achieves a
steady value and the resulting $\alpha_M$ is large enough
to offset the turbulent diffusion. Then the large scale field
decays initially at rate of the order resistive rate.
When $M_0$ is below a critical value,
the initial helicity in large scales is insufficient 
to generate a large enough small scale
helicity, and $\alpha_M$, by turbulent diffusion.
In this case the SSH decays, and the LSHE decays
fast due to turbulent diffusion (uncompensated by the $\alpha_M$ effect).

It can be seen from the right panel of 
Fig.~\ref{figvaryinit} that the dash-dotted blue line 
starting with $M_0 = 0.05$ is the last to reach a steady state
indicating the presence of resistive decay regime initially.
For such smaller $M_0 \sim 0.04$ and below,
the small scale helicity fails to rise to a steady state and
subsequently decays indicating the absence of slow decay 
regime, which means that the large scale field directly
starts decaying at a faster rate.
This can be seen from the left panel of Fig.~\ref{figvaryinit}.
Note that curves with $M_0$ above the threshold
have sharp initial drop (as SSH builds up), but do not decrease in their energy significantly
before joining the fiducial curve.
On the other hand, the time-shifted purple dotted curve with $M_0=0.02$, 
which is below the threshold, drops by several orders of magnitude before joining the fiducial curve.
Hence, we find that the transition point
is close to the value of $(E_{c2}/M_{eq})=(k_1/\kf)^2=0.04$
as expected from the work of Paper I and the Eq.~\ref{Meq} above. 

Again the same exercise is repeated at $\kf=20$. In the Fig.~\ref{figvaryinitkf20},  
the solid green curve starting with $M_0=0.001$,
which is below the expected critical energy of $(k_1/\kf)^2$,
is seen to drop in energy by few orders of magnitude before joining the fiducial curve.
Hence in this case, we find the transition
energy, $E_{c2} \sim 0.003 \ M_{eq}$, as is seen from the Fig.~\ref{figvaryinitkf20}.
This value of $E_{c2}$ for the is again close
to $(k_1/\kf)^2 M_{eq}$.
Such large $\Rm=12000$ is however beyond the scope of current DNS. Thus in the
DNS studies below, where $\Rm$ has a more modest value of $\sim 100$, we
may not expect to see such a clear evidence of $E_{c2}$.
Nevertheless, we do expect to check the consistency with the two
scale model and hence indirectly substantiate the results of Paper I.
\begin{figure}
\epsfig{file=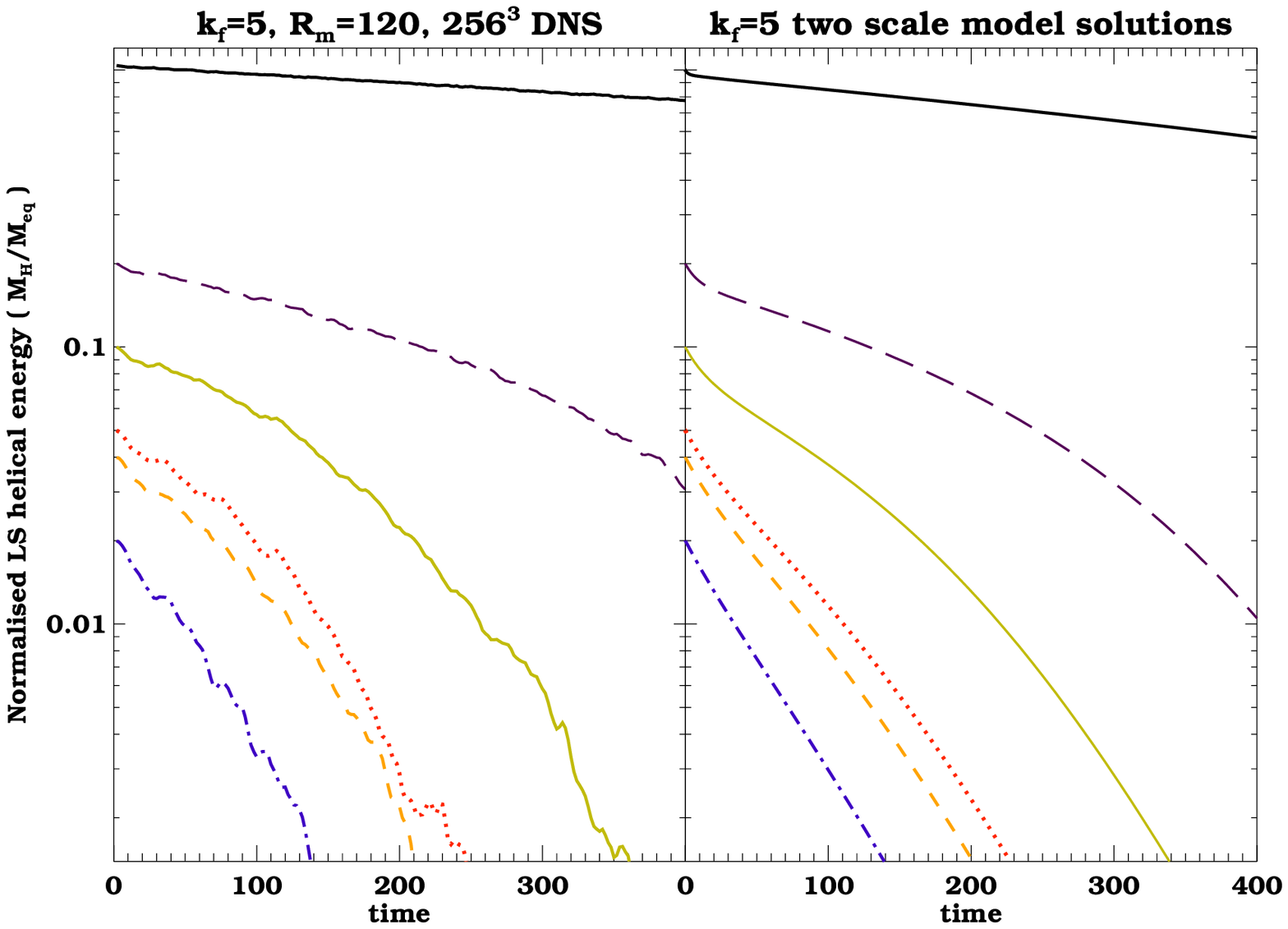, width=0.475\textwidth, height=0.3\textheight}
\epsfig{file=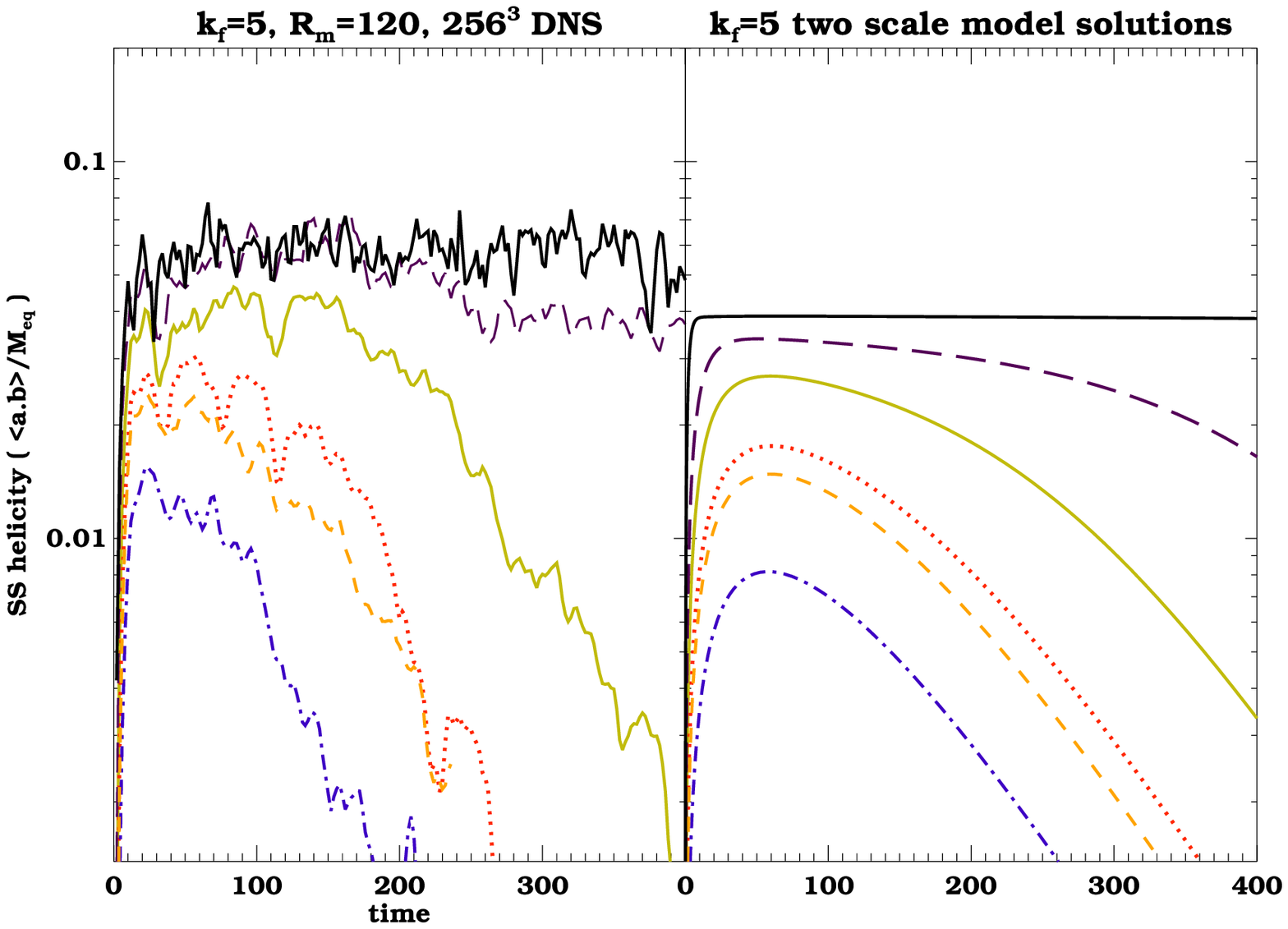, width=0.475\textwidth, height=0.3\textheight}
\caption{The decay curves for fully helical large scale magnetic field with 
different $M_0$ for $\kf=5$ and $\Rm=120$, from DNS, are shown.}
\label{figsimvary}
\end{figure}
\begin{figure}
\epsfig{file=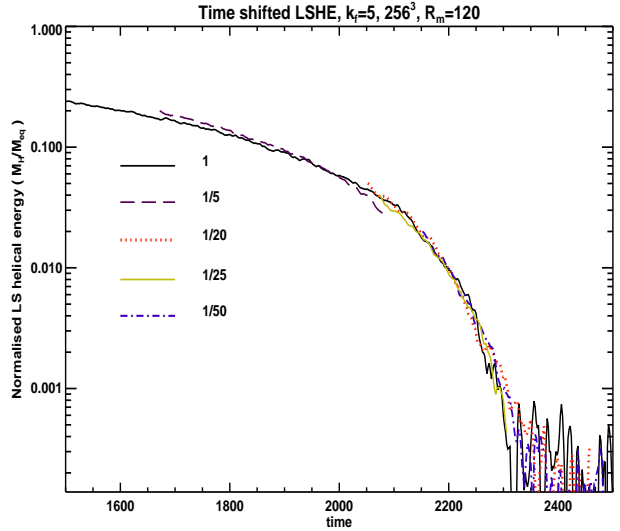, width=0.475\textwidth, height=0.3\textheight}
\caption{The time-shifted decay curve for fully helical large scale magnetic field for 
different sub-equipartition $M_0$ at $R_M\simeq 120$. This value of $R_M$ does not meet the
condition $3/R_M \ll (k_1/k_f)^2$ needed to identify the predicted transition (see text).}
\label{figshiftsim}
\end{figure}

We have shown in Fig.~\ref{figsimvary}, the results from the DNS runs
G-K where the initial magnetic
energy is lowered to smaller and smaller values compared to $M_{eq}$.
In these set of simulations, we have fixed $k_f=5$, $u_{rms} \sim 0.12$ 
and $\eta=2 \times 10^{-4}$, and thus $\Rm = 120$, while varying $M_0$.
The top left panel of Fig.~\ref{figsimvary} shows the time evolution of large scale 
magnetic energy in the DNS, while the top right panel the results from the
corresponding two scale model. The time evolution of the SSH 
in the DNS and corresponding two scale model are respectively 
shown in the bottom left and right panels of Fig.~\ref{figsimvary}.

A comparison beween the DNS (left panel) and the 2-scale model (right panel) 
in Fig.~\ref{figsimvary}, shows that there is a qualitative agreement between the two.
For example, the slow decay rate of large scale energy of both are comparable, and
so also are the amplitudes of the steady state small scale helicity.
We also find that the average initial slopes (evaluated for the time period
of t=0 to t=100) for all the runs G-K, listed in the Table~\ref{xxx}, match
closely with the estimate using Eq.~\ref{relationJBjb}, where the term
in the denominatior, $\rmsB/2{M_{eq}}$ is replaced with the value of $M_0$.

In Fig.~\ref{figshiftsim}, we show the fiducial curve where 
$M_0=1$ is the solid 
black line.
We also show the evolution curves starting with 
sub-equipartition energies, time-shifted to lie on the fiducial curve.
As pointed out earlier, a clear energy transition value would be
revealed if curves with $M_0$ below the threshold
decrease in their energy significantly
before joining the fiducial curve.
The graph shows  all the curves 
nearly falling together without any such drop in the initial
energy which at face value means no clear identification of $E_{c2}$
in the energy scale.
However, from Eq.~\ref{relationJBjb},
it can be seen that the term $\eta/\eta_t=3/\Rm$ has to be sufficiently 
small  compared to $M_0$,
to be able to discern the $E_{c2}$
dependence on $\kf$, for the subsequent evolution.
If we compare the two terms in the denominator of Eq.~\ref{relationJBjb}, then we require, $3/\Rm << (k_1/\kf)^2$.
Otherwise, $\eta/\eta_t$ would become important before
$M_0$ is lowered to $(k_1/\kf)^2$, and one cannot discern the influence of 
$\kf$ on
$E_{c2}$.
With $\kf=5$ and $\Rm=120$, $3/\Rm=0.025$ and $(k_1/\kf)^2=0.04$,
thus the two terms are comparable. Hence, we seemingly need much higher $\Rm$ runs
\footnote{On the other hand, in the context of $E_{c1}$, it is the term $\eta/\eta_t$ 
which is responsible for the transition to faster decay phase.
And we have seen that the transition energy $E_{c1}$ depends on $\Rm$, scaling
as $\Rm^{-1/2}$ from the two scale model solutions. 
Thus DNS with even a modest $\Rm$ can enable us to
discern the transition energy, $E_{c1}$, which morover agrees 
reasonably with that predicted by
the corresponding 
two scale model solutions.}
to be able to properly check the more conservative threshold
of $E_{c2}/M_{eq}=(k_1/\kf)^2$.

Note that in  Fig.~\ref{figsimvary}, as $M_0$
is decreased,  approaching $E_{c2}$ from above,
 the subsequent decay seems to be at an increasingly higher rate.
This can be  understood from Eq.~\ref{relationJBjb} (which applies only for $M_0> E_{c2}$ since only in that regime does $\fluccurhel$ 
reach the steady state assumed by that equation.)
For the curves with lower initial energy, the ratio, $\fluccurhel/\meancurhel$
will be larger, thus leading to a higher multiple of the resistive decay rate.
This can be understood  from examining Fig.~\ref{figshiftsim}.
Since the curves with lower $M_0$, fall on the
curve with $M_0=1$, later in time within the slow decay phase, they are
expected to decay at an increasingly larger rate.

Overall we see a qualitative agreement between the DNS
and the 2-scale model. This again indicates that the result of Paper I
on the slow decay of helical magnetic fields which have 
$M_0 > (k_1/k_f)^2 M_{eq}$, seems reasonably
consistent with the simulations that we have performed so far.
Though higher resolution simulations with high $\Rm$
and small sub-equipartition initial fields are required to substantiate
the above results.
\subsection{Role of SSH in explaining $E_{c1}\ne E_{c2}$}

In the distinct contexts of the previous two sections where we have identified the
transition energies  $E_{c1}$ and $E_{c2}$ we started with the SSH initially equal to zero.
Here we discuss how the distinction between $E_{c1}$ and $E_{c2}$ can be traced to
the distinct levels to which $\fluchel$  builds up in the two cases.

The first context (section 3.1)   in which we have identified the 
the transition energy
is 
the fiducial case of $M_0=1$,
for which we found a $\kf$ independent transition point $E_{c1}$.
In this case,
$\fluchel$ builds up to a nearly steady state value during much of the resistive slow decay  phase of $\meanhel$  before the
fast decay occurs.   
In contrast, for the  second context  (section 3.2) of varying $M_0$,
we found that when fast decay of the large scale field occurs right from the beginning, 
 the helicity transferred from large scale to small scales
 never attains the aforementioned steady state value.
The fast decay happens below a critical initial large scale helical energy value $E_{c2}$.
If there is not enough initial helical large scale energy to supply the needed SSH,
the large scale field decays fast. A source of SSH  is crucial to explain the threshold of $E_{c2}$ on $\kf$ (Paper I).

To quantitatively study the  importance of the role of  SSH source in distinguishing  $E_{c1}$ and $E_{c2}$,
we  can ask if setting the initial SSH equal to the maximum
steady state value of the case of section 3.1 
(rather than allow it to grow from zero)  and then vary the initial $M_0$ (as in the case of section 3.2)
do we recover  $E_{c1}$?

\begin{figure}
\epsfig{file=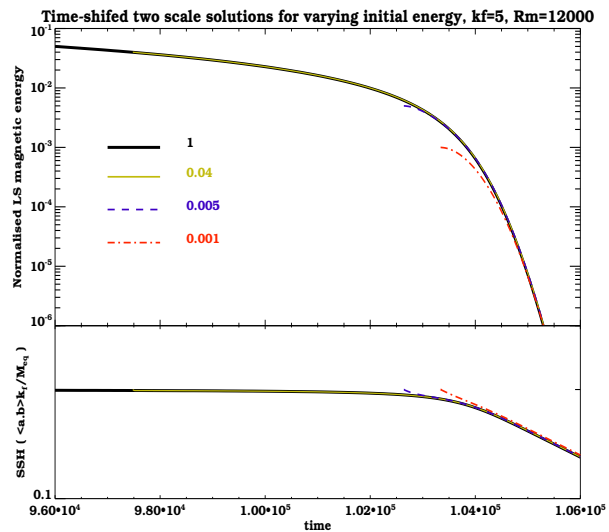, width=0.475\textwidth, height=0.3\textheight}
\caption{The two scale solutions for fully helical large scale magnetic field (in top panel)
and SSH (in bottom panel) with  
different $M_0$ for $\kf=5$ and $\Rm=12000$, where the initial SSH $\ne$ 0, but 
is set to the value of $(k_1/\kf^2)M_{eq}$}
\label{figh2init}
\end{figure}

\begin{figure}
\epsfig{file=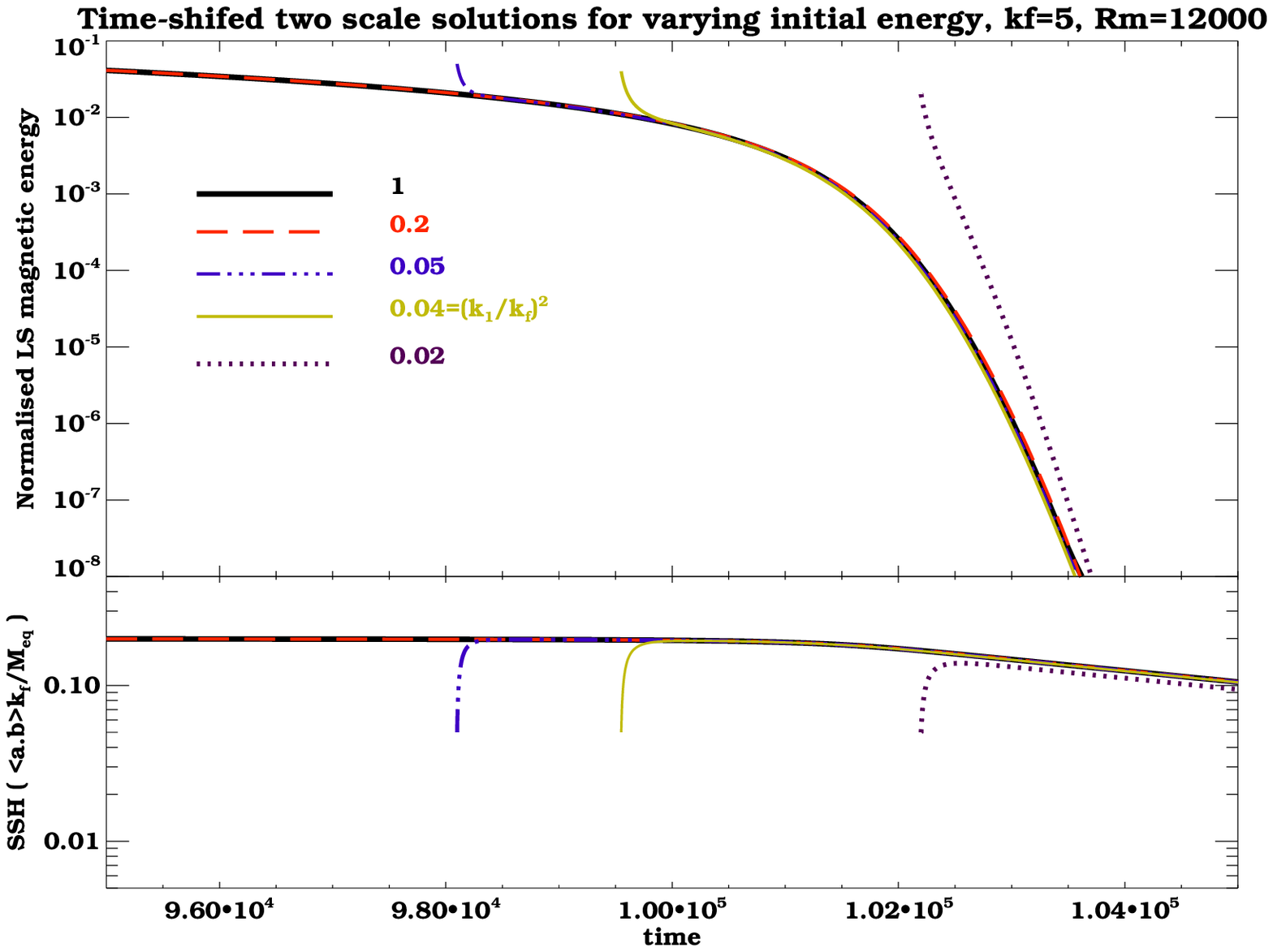, width=0.475\textwidth, height=0.3\textheight}
\caption{The two scale solutions for fully helical large scale magnetic field (in top panel)
and SSH (in bottom panel) with  
different $M_0$ for $\kf=5$ and $\Rm=12000$, where the initial SSH $\ne$ 0, but 
is set to the value of $\fluchel \kf/M_{eq}=0.2$}
\label{figh2init05}
\end{figure}

Indeed, it is seen from Fig.~\ref{figh2init}, in the case of $\kf=5$,
$\Rm=12000$ 
and with the initial $\fluchel=(k_1/\kf^2)M_{eq}$, the two scale model solutions
for varying $M_0$ fall together on the fiducial evolution curve for the context in section 3.1
For evolution starting at lower and lower energy values, the LSHE starts off
initially with a flatter slope, but eventually joins the fiducial
curve, which at that point in time is decaying at a much larger rate.
The corresponding SSH quickly decays from the initial value of $(k_1/\kf^2) M_{eq}$
to the value on the fiducial curve at that point in time.
Additionally, 
it can be seen from Fig.~\ref{figh2init05}
that if SSH is set to a value, $0<\fluchel<(k_1/\kf^2)M_{eq}$,
then the two scale model solutions are similar to the second context, where SSH
initially rises to attain the steady state value, and fails to do so
when $M_0$ is at the threshold or below.
This shows that the difference between the two contexts that led to 
 $E_{c1}\ne E_{c2}$  obtains due to the difference in the value of SSH attained in the early transient
phase of the evolution.

\section{Discussion and Conclusions}

The extent to which large scale fields survive turbulent diffusion in the absence of dynamo action via kinetic helicity or shear
is important in assessing the plausibility of dynamo versus fossil field origin of large scale fields in astrophysical objects.
Non-helical fields decay at the turbulent diffusion rate in  the presence of non-helical turbulence, but 
large scale
helical fields do not 
(Paper I).
Here we have examined the survival of initially helical fields via direct numerical simulations (DNS),
and  compared the results with the basic two scale model of Paper I. 
Previous simulations have been done by \citet{YBG03} and \citet{KBJ11}.
In particular, we have examined the decay of 
large scale
helical fields in  more general settings by studying in detail the dependence on initial 
field strength, forcing wavenumber of the turbulence and also $R_M$.
DNS takes into account the full set of MHD equations but 
is limited by computational power to modest Rm of order 100.
On the other hand, the two scale model involves simplifying assumptions, of having only 
two scales and also invokes a closure approximation for the turbulent emf, 
but allows large Rm to be explored.
Thus comparison will allow one to evaluate to what extent
the two scale model can be trusted, and also then the possibility
of extrapolating DNS to larger Rm.
Overall, we find that there is good qualitative agreement between the predictions of two-scale theory and DNS.

For the case in which we start with an initial helical magnetic energy of sufficient strength,
of order the equipartition value ($M_0 \sim 1$),
the fields first exhibits a slow resistive decay.  The decay rate  steadily increases until the 
large scale helical energy falls to a few percent (3-5\%) 
of the equipartition value.   The resistive decay phase matches the predictions of the two scale model quite well.
We can closely match the decay slope in this phase with Eq. 14. 
Moreover, we show that the assumption that $\alpha_K$ is much smaller
than $\alpha_M$, which was made in Paper I, holds true.
This indicates that the basic picture of Paper I,
particularly regarding the importance of a magnetic
alpha, $\alpha_M$, being generated by turbulent diffusion of the large scale
helical field, and then itself acting to prevent the turbulent
decay of the large scale field, is reasonably robust. This also
indirectly supports the ideas behind the closure approximations
used to derive the magnetic alpha. 

Subsequently, there is a transition to  fast decay phase, which can
exceed even the turbulent decay rate predicted by a two scale theory.
For this fiducial case of starting with equipartition energy, 
the threshold energy at which the transition occurs, $E_{c1}$,  
is independent of the forcing wavenumber $k_f$ and 
$\Rm$ for the range of $\Rm \sim 100$ explored. 
Meanwhile, the 2-scale model solutions at a much higher Rm=12000, do indicate a
possible scaling of the transition energy $E_{c1} \propto \Rm^{-1/2}$.

A different transition energy threshold $E_{c2}$ arises
for the case in which we seek the transition threshold at $t=0$
below which the field decays at a fast rate right from the beginning.
This scenario is more astrophysically relevant, since the
feasibility of the existence of subequipartition strength initial fields is 
higher as compared to the fiducial case.
In such a case,  
Paper I argues that when the large scale
helical field energy is
below a critical initial magnetic energy $E_{c2} \sim (k_1/k_f)^2 M_{eq}$,
it decays rapidly at the turbulent diffusion rate.
We have reconfirmed this estimate by solving the two
scale model exactly in the ideal limit, and also solving it numerically for finite but very large $\Rm$,
much larger than possible by DNS.
For the moderate $\Rm \sim 100$ achievable by DNS (for DNS with large $\kf$), we have shown again
that the DNS results are consistent with the 2-scale model.
However, robustly identifying the transition energy $E_{c2}$  predicted by the 2-scale model,
requires the condition $3/R_M << (k_1/k_f)^2$ to be satisfied, and $\Rm$ is 
not sufficiently large 
in the simulations.
Much  higher $\Rm$ simulations would be required. At present we can only say that,
from the overall consistency between DNS and 2-scale model even for case 1, 
we do expect this later type of transition to obtain for high $R_M$ cases.

 Eventually it would be desirable to assess how the principles identified herein apply to more realistic conditions of astrophysical rotators to assess whether large scale fields in astrophysical rotators such as galaxies could result from post-processing of fossil helical fields without requiring a traditional in situ kinetic helicity.
Real systems  have shear, differential rotation, and stratification, all of which we have not considered here.   We also  considered all large scale quantities to be averaged over closed volumes in the present work, thereby eliminating helicity fluxes.  It would be of interest for future work to consider the influence of helicity fluxes on the relative decay of helical and non-helical large scale fields.  Finally we note that in real systems,  there would in general be a combination of helical and non-helical fields and the results herein would  apply  to the helical fraction of the large scale magnetic energy.

\section*{Acknowledgments}
We thank Axel Brandenburg for useful discussions
which helped to sharpen the arguments of this paper.
We acknowledge the use of the high performance computing facility at IUCAA.

%---------------------------------------------
\bibliographystyle{mn2e}
\bibliography{decayrefs}

\label{lastpage}

\end{document}